\def\ds{\displaystyle}
\def\overc{\overline c}
\def\overb{\overline b}
\def\overt{\overline t}
\def\overq{\overline q}

\def\fakebold#1{\leavevmode\setbox0=\hbox{#1}%
  \kern-.025em\copy0 \kern-\wd0
  \kern .05em\copy0 \kern-\wd0
  \kern-.025em\raise.0433em\box0
}

\def\epslash{\epsilon\kern-0.47em/}
\def\Aslash{A\kern-0.47em/}
\magnification=\magstephalf \baselineskip=12pt \centerline {\bf A
Variational Fock-Space Treatment of Quarkonium}
\par\medskip
\centerline {L. Di Leo and J.W. Darewych} \centerline {Department
of Physics and Astronomy} \centerline {York University}
\centerline {Toronto, ON} \centerline {M3J 1P3 Canada}
\par\medskip
\par\medskip
\par\medskip
\par\medskip
\par\medskip
\par\medskip
\par\medskip
\par\medskip
\par\medskip
\par\medskip
\par\medskip
\centerline {Abstract}
\par\medskip
The variational method and the Hamiltonian formalism of QCD are
used to derive relativistic, momentum space integral equations for
a quark-antiquark system with an arbitrary number of gluons
present. As a first step, the resulting infinite chain of coupled
equations is solved in the nonrelativistic limit by an approximate
decoupling method. Comparison with experiment allows us to fix the
quark mass and coupling constant, allowing for the calculation of
the spectra of massive systems such as charmonium and bottomonium.
Studying the results with and without the nonAbelian terms, we
find that the presence of the nonAbelian factors yields better
agreement with the experimental spectra.
\par\medskip
\par\medskip
\par\medskip
\par\medskip
\par\medskip
\par\medskip
\par\medskip
\par\medskip
\par\medskip
\par\medskip
\par\medskip
\par\medskip
\par\medskip
\par\medskip
\par\medskip
PACS number(s): 11.10.Ef,12.38.-t,14.40.-n
\par\vfill\eject
\centerline {I. INTRODUCTION}
\par\medskip
Although the hadron spectrum is understood quite well in the
context of the quark model, completely {\it ab initio} treatments
have proved to be difficult to implement, particularly outside the
realm of lattice gauge theory. In this paper we shall consider the
description of quark-antiquark states using a variational approach
within the canonical Hamiltonian formalism of QCD. This approach
has been used with good effect for describing relativistic two and
three body bound and quasi-bound states in various QFTs, including
QED [1-6], the Wick-Cutkosky model [7], model QCD [8], and other
models. An overview of the variational approach in QFT up to 1988
is given in the conference proceedings mentioned in [1]. A brief
review of the description of few-particle bound and quasi-bound
states in QFT by means of the variational approach in the
Hamiltonian formalism is given in reference [9].
\par
The present work is similar, in some respects, to the approach
used by Zhang and Koniuk [8], except that they used a model
Hamiltonian in which a scalar confining potential is put in by
hand. In the present work we use the full QCD Hamiltonian in the
Lorentz gauge as explained in Section II. The variational trial
state for the quarkonium system is a Fock-space expansion that
includes terms of the type $|\phi_0\;q\;\overq\rangle$,
$|\phi_1\;q\;\overq\;g\rangle$, $|\phi_2\;q\;\overq\;g\;g\rangle$,
where the $\phi_{\beta}$ are variational coefficient functions and
$q$, $\overq$ and $g$ represent quarks, antiquarks and gluons
respectively. We use a Gupta-Bleuler type constraint to implement
the Lorentz gauge condition.
\par
The variational principle then leads to an infinite chain of
coupled integral equations for the functions $\phi_{\beta}$, which
are given in Section III. Such a system of equations is impossible
to solve, so the rest of the paper deals with some approximate
solutions. Thus, in Section IV we consider the limit of heavy
quark masses, and approximate solutions of the nonrelativistic
equations are discussed in Section V, as would be adequate for the
heavy $c\overc$ and $b\overb$ systems. A comparison with some
observed charmonium and bottomonium states is presented and
discussed. Concluding remarks are given in Section VI.
\par\medskip
\centerline {II. LAGRANGIAN, HAMILTONIAN AND CANONICAL
QUANTIZATION}
\par\medskip
Suppressing the flavor indices, the QCD Lagrangian density is [11]
$$ {\cal L}_{QCD}=-{1\over
4}F^{a\;\mu\,\nu}\;F_{\mu\,\nu}^a+i\;{\bar
\psi}^i\;\gamma^{\mu}\;(D_{\mu})_{i\,j}\;\psi^j-m\;{\bar
\psi}^i\;\psi^i-(m_0-m)\;{\bar \psi}^i\;\psi^i \eqno (1) $$ where
$$
F^a_{\mu\,\nu}=\partial_{\mu}\,A^a_{\nu}-\partial_{\nu}\,A^a_{\mu
}+ g_s\;f_{a\,b\,c}\;A^b_{\mu}\;A^c_{\nu}, \eqno (2) $$ and $$
(D_{\mu})_{i\,j}=\delta_{i\,j}\;\partial_{\mu}-i\;g_s\;t^a_{i\,j}
\;A^a_{\mu}. \eqno (3) $$ The QCD coupling constant is denoted as
$g_s$, $f_{a\,b\,c}$ are the structure constants, the fermion mass
$m$ is an adjustable parameter not necessarily equivalent to the
bare fermion mass $m_0$, and $$ t^a_{i\,j}={\lambda^a_{i\,j}\over
2}, \eqno (4) $$ where the $\lambda$s are the Gell-Mann matrices.
As usual, repeated indices are summed over, with the color indices
$i,j=1,2,3$ and $a,b,c=1,...,8$ for the gluon fields.
\par
Upon expansion of the terms above, the Lagrangian density may also
be written as $$ {\cal L}_{QCD}={\cal L}_{\psi}+{\cal L}_{g}+{\cal
L}_{g\;\psi}+{\cal L}_{3\,g}+{\cal L}_{4\,g} \eqno (5) $$ where
$$\eqalignno{ &{\cal L}_{\psi}={\bar
\psi}^i\;\left(i\;\gamma^{\mu}\;\partial_{\mu}-m\right)\psi^i
-(m_0-m)\;{\bar \psi}^i\;\psi^i &(6)\cr &{\cal L}_{g}=-{1\over
2}\;\partial_{\mu}\,A^a_{\nu}(\partial^{\mu}\,A^{a\;\nu} -
\partial^{\nu}\,A^{a\;\mu}) &(7)\cr &{\cal
L}_{g\,\psi}=g_s\;t^a_{i\,j}\;{\bar \psi}^i\;\;{\Aslash}^a\psi^j
&(8)\cr &{\cal
L}_{3\,g}=-\partial^{\mu}\,A^{a\;\nu}\;g_s\;f_{a\,b\,c}\;
A^{b}_{\mu}\;A^{c}_{\nu} &(9)\cr &{\cal L}_{4\,g}=-{1\over
4}\;g_s^2\;f_{a\,b\,c}\;f_{a\,d\,e}\;A^{b}_{\mu}\;A^{c}_{\nu}\;A^
{d\;\mu}\;A^{e\;\nu} &(10)\cr} $$
\par
Since the Lagrangian yields $$ \Pi^0_a(x)={\partial {\cal
L}_{QCD}\over \partial {\dot A}^a_0}=0 \eqno (11) $$ we cannot
quantize unless we use the Lorentz condition $$
\partial_{\mu}\;A^{a\;\mu}=0 \eqno (12)
$$ and add a gauge fixing term [12] to $\ds {\cal L}_g$ such that
$$ {\cal L}_g\rightarrow -{1\over
2}\;\partial_{\mu}\,A^a_{\nu}\;\partial^{\mu}\,A^{a\;\nu} \eqno
(13) $$ We will use in the quantized theory, in place of (12), the
weaker Gupta-Bleuler (GB) constraint [12] $$
\partial_{\mu}\;{A^{a\;\mu}}^+(x)|\psi\rangle =0.
$$
\par
The Hamiltonian density is canonically derived from $$ {\cal
H}_{QCD}=\Pi^k\;{\dot \psi}^k+ \Pi^{\alpha}_a\;{\dot
A}^a_{\alpha}-{\cal L}_{QCD} \eqno(14) $$ where $$ \Pi^k={\partial
\,{\cal L}_{QCD}\over \partial \, {\dot \psi}^k}={\bar
\psi}^i\;i\;\gamma^0\;\delta_{i\;k},\;\;\;\;
\Pi^{a\;\alpha}={\partial \,{\cal L}_{QCD}\over \partial \, {\dot
A}_{\alpha}^a}=-{\dot
A}^{a\;\alpha}-g_s\;f_{a\,b\,c}\;A^b_0\;A^{c\;\alpha} \eqno (15)
$$ so that using (5) we can write $$ {\cal H}_{QCD}={\cal
H}_{\psi}+{\cal H}_{g}+{\cal H}_{\psi\;g}+{\cal H}_{3g}+{\cal
H}_{4g}\eqno (16) $$ where $$ {\cal H}_{\psi}={\bar
\psi}^i\;(-i\;\gamma^l\;\partial_l+m)\psi^i+(m_0-m){\bar
\psi}^i\;\psi^i, \eqno (17) $$ $$ {\cal H}_g=-{1\over
2}\;\Pi^a_{\nu}\;\Pi^{a\;\nu}-{1\over
2}\;\partial_l\;A^a_{\nu}\;\partial^l\;A^{a\;\nu}, \eqno (18) $$
$$ {\cal H}_{\psi\;g}=-g_s\;t^a_{i\;j}\;{\bar
\psi}^i\;\Aslash^a\;\psi^j, \eqno (19) $$ $$ {\cal H}_{3g}=-
g_s\;f_{a\,b\,c}\;\Pi^{a\;\nu}\;A^b_0\;A^c_{\nu}+g_s\;f_{a\,b\,c}
\;\partial^l\;A^{a\;\nu}\;A^b_l\;A^c_{\nu}, \eqno (20) $$ and $$
{\cal H}_{4g}=- {1\over
4}\;g_s^2\;f_{a\,b\,c}\;f_{a\,d\,e}\;A^c_{\nu}\;A^{e\;\nu}\;
\left[ A^b_{0}\;A^{d\;0}+A^b_{l}\;A^{d\;l} \right] . \eqno (21) $$
\par
We can expand the Dirac field ($t=0$, color index $j$) and gluon
field ($t=0$, gluon index $a$) in the standard Fourier
decomposition: $$ \psi^j({\bf x})=\sum_s\int d^3p \sqrt{{m\over
(2\,\pi)^3\;\omega_p}}\left[c^j({\bf p},s)\;u({\bf
p},s)\;e^{i\,{\bf p}\cdot{\bf x}}+d^{j\;\dagger}({\bf
p},s)\;v({\bf p},s)\;e^{-i\,{\bf p}\cdot{\bf x}}\right], \eqno
(22) $$ $$ {\bar \psi^j}({\bf x})=\sum_s\int d^3p \sqrt{{m\over
(2\,\pi)^3\;\omega_p}}\left[d^{j}({\bf p},s)\;{\bar v}({\bf
p},s)\;e^{i\,{\bf p}\cdot{\bf x}}+ c^{j\;\dagger}({\bf
p},s)\;{\bar u}({\bf p},s)\;e^{-i\,{\bf p}\cdot{\bf x}}\right],
\eqno (23) $$ $$ A^{a\;\mu}({\bf x})=\sum_{\lambda=0}^{3}\int
{d^3k\over \sqrt{(2\;\pi)^3\;2\;| {\bf k}
|}}\epsilon^{\mu}_{\lambda}({\bf k})\left[ a^a_{\lambda}( {\bf
k})\;e^{i\,{\bf k}\cdot{\bf x}}+ a^{a\;\dagger}_{\lambda}({\bf
k})\;e^{-i\,{\bf k}\cdot{\bf x}}\right], \eqno (24) $$ and $$
\Pi^{a \; \nu}({\bf x})=i\;\sum_{\lambda=0}^{3}\int {d^3k\;|{\bf
k}|\over \sqrt{(2\;\pi)^3\;2\;|{\bf
k}|}}\epsilon^{\nu}_{\lambda}({\bf k})\left[ a^a_{\lambda}({\bf
k})\;e^{i\,{\bf k}\cdot{\bf x}}-a^{a\;\dagger}_{\lambda}({\bf
k})\;e^{-i\,{\bf k}\cdot{\bf x}}\right], \eqno (25) $$ where
$c^{j\;\dagger}$, $d^{j\;\dagger}$ and $a^{a\;\dagger}_{\lambda}$
are the creation operators of a (free) fermion, antifermion and
gluon respectively, and $\omega_p=\sqrt{{\bf p}^2+m^2}$.
Furthermore, the anticommutation and commutation relations are $$
\{c^i({\bf p},s),c^{j\;\dagger}({\bf p}',s')\}=\{d^i({\bf
p},s),d^{j\;\dagger}({\bf
p}',s')\}=\delta_{i\;j}\;\delta_{s\;s'}\;\delta^3({\bf p}-{\bf
p}') \eqno (26) $$ $$ [a^a_{\lambda}({\bf
k}),a^{b\;\dagger}_{\lambda'}({\bf
k}')]=\zeta_{\lambda}\;\delta_{a\;b}\;\delta_{\lambda\;\lambda'}
\;\delta^3({\bf k}-{\bf k}') \eqno (27) $$ where
$\zeta_1=\zeta_2=\zeta_3=1$, $\zeta_0=-1$, and all other
commutation relations vanish.
\par
Since we are not interested in the vacuum energy, we will
normal-order the Hamiltonian $$ :H_{QCD}:=\int d^3x :{\cal
H}_{QCD}: \eqno (28) $$ and expand its component parts using
equations (17) to (25) in (28).
\par\medskip
\centerline {III. COUPLED EQUATIONS}
\par\medskip
For a quark-antiquark system and an arbitrary number of gluons, we
will use a Fock-space variational ansatz (rest frame ${\bf
P}_{total}=0$) that is a linear combination of a quark-antiquark
state, a quark-antiquark-gluon state, a
quark-antiquark-gluon-gluon state, {\it ad infinitum}. Without the
terms that contain Fock states with at least two gluons, we would
not be able to sample the nonAbelian terms of the Hamiltonian.
Explicitly, $$ \eqalignno{ &|\psi\rangle=\sum_{i,s_1,s_2}\;\int
d^3p\;d^3n\;\phi_0({\bf p},{\bf n},s_1,s_2)\;c^{i\;\dagger}({\bf
p},s_1)\;d^{i\;\dagger}({\bf n},s_2)\;\delta^3({\bf p}+{\bf
n})|0\rangle\cr &+\sum_{i,j,s_1,s_2}\;\sum_{z_1,\lambda_1}\;\int
d^3p\;d^3n\;d^3q_1\;\phi_1({\bf p},{\bf n},s_1,s_2,{\bf
q}_1,\lambda_1)\;c^{i\;\dagger}({\bf
p},s_1)\;t^{z_1}_{i\,j}\;d^{j\;\dagger}({\bf
n},s_2)\;a^{z_1\;\dagger}_{\lambda_1}({\bf q}_1)\cr &\times
\delta^3({\bf p}+{\bf n}+{\bf q}_1)|0\rangle \cr
&+\sum_{i,j,l_1,s_1,s_2}\;\sum_{z_1,z_2,\lambda_1,\lambda_2}\;
\int d^3p\;d^3n\;d^3q_1\;d^3q_2\;\phi_2({\bf p},{\bf
n},s_1,s_2,{\bf q}_1,{\bf
q}_2,\lambda_1,\lambda_2)\;c^{i\;\dagger}({\bf p},s_1)\cr &\times
(t^{z_1}_{i\;l_1}\;t^{z_2}_{l_1\;j}+t^{z_2}_{i\;l_1}\;t^{z_1}_{l_
1\;j})\;d^{j\;\dagger}({\bf
n},s_2)\;a^{z_1\;\dagger}_{\lambda_1}({\bf
q}_1)\;a^{z_2\;\dagger}_{\lambda_2}({\bf q}_2)\;\delta^3({\bf
p}+{\bf n}+{\bf q}_1+{\bf q}_2)|0\rangle+...\cr &=
\sum_{s_1,s_2}\sum_{i,j}^3 \int d^3p\;d^3n \;c^{i\;\dag}({\bf
p},s_1)\;d^{j\;\dag}({\bf n},s_2)\cr
&\times\sum_{\beta=0}^{\infty}\;\sum_{\lambda_1, \ldots
,\lambda_{\beta}}^{3}\; \sum_{l_1,\ldots,
{l_{\beta+1}}}^3\;\sum_{z_1, \ldots, z_{\beta}}^8 \int
d^3q_1\cdots d^3q_\beta\;\phi_{\beta} ({\bf p},{\bf
n},s_1,s_2,q,\lambda)\;\delta_{l_1\,i}\delta_{l_{\beta+1}\,j} \cr
&\times \{t^{z_1}_{l_1\;l_2}\;t^{z_2}_{l_2\;l_3}\cdots
t^{z_{\beta}}_{l_\beta\;{l_{\beta+1}}}\}\;
a^{z_1\;\dag}_{\lambda_1}({\bf
q}_1)\;a^{z_2\;\dag}_{\lambda_2}({\bf q}_2)\cdots
a^{z_{\beta}\;\dag}_{\lambda_{\beta}}({\bf
q}_\beta)\;\delta^3({\bf p}+{\bf n}+{\bf q}_1+\cdots+{\bf
q}_\beta)|0\rangle, & (29)\cr} $$ where the $\lambda_k$'s are
summed from $0$ to $3$, $i$, $j$, $k$, $l$, $m$ and the $l_k$'s
from $1$ to $3$, and the $z_k$'s from $1$ to $8$, and the
$\{\cdots\}$ represent all ($\beta!$) permutations of the
Gell-Mann matrices.  The total momentum is zero in the rest frame
and the annihilation operators $c^j$, $d^j$ and $a^a_{\lambda}$
have the property $$
c^j|0\rangle=d^j|0\rangle=a^a_{\lambda}|0\rangle=0, $$ where
$|0\rangle$ is our trial vacuum state. In our ansatz, we have used
the simplified notation [13] $$ q={\bf q}_1,\ldots,{\bf
q}_\beta,\;\;\; \lambda=\lambda_1,\ldots,\lambda_\beta $$ and
define $\tilde q_\beta$ as the set of all the $q$'s excluding
$q_\beta$. We will use this same definition for the $z_k$'s and
$t^{z_k}$'s in what follows. The symmetry of $\phi_{\beta}({\bf
p},{\bf n},s_1,s_2,{\bf q}_1,\ldots,{\bf
q}_\beta,\lambda_1,\ldots,\lambda_\beta)$ is apparent under the
exchange of any $({\bf q}_k,\lambda_k)$ pair.
\par
The GB condition removes the redundant gluon degrees of freedom by
requiring that our trial state $|\psi\rangle$ obeys the condition
$$ \Biggl(a^{z_1}_{\lambda=3}({\bf q}_1)-a^{z_1}_{\lambda=0}({\bf
q}_1)\Biggr)|\psi\rangle=0 $$ for all ${\bf q}_1$, which forces
the variational coefficients $\phi_{\beta}$ to obey $$\eqalignno{
&\phi_1({\bf p},{\bf n},s_1,s_2,{\bf
q}_1,\lambda_1=3)=-\phi_1({\bf p},{\bf n},s_1,s_2,{\bf
q}_1,\lambda_1=0)\cr &\phi_2({\bf p},{\bf n},s_1,s_2,{\bf
q}_1,{\bf q}_2,\lambda_1,\lambda_2=3)=-\phi_2({\bf p},{\bf
n},s_1,s_2,{\bf q}_1,{\bf q}_2,\lambda_1,\lambda_2=0)\cr
&\phi_2({\bf p},{\bf n},s_1,s_2,{\bf q}_1,{\bf
q}_2,\lambda_1=3,\lambda_2)=-\phi_2({\bf p},{\bf n},s_1,s_2,{\bf
q}_1,{\bf q}_2,\lambda_1=0,\lambda_2)\cr} $$ etc.
\par
The variational coefficients $\phi_0$, $\phi_1$,...$\phi_{\beta}$
in (29) are determined from the variational principle $$
\langle\delta\;\psi|:H_{QCD}-E:|\psi\rangle=0 \eqno (30) $$ which,
with our trial state and the GB condition, gives the infinite
chain of coupled, multidimensional integral equations:
$$\eqalignno{ &\left[
\omega_{p}+\omega_{n}+m\;(m_0-m)\left({1\over \omega_{p}}+ {1\over
\omega_{n}}\right)+\sum_{k=1}^{\beta}\;|{\bf q}_k|-E
\right]\;I^{\beta}\;\phi_{\beta} ({\bf p},{\bf
n},s_1,s_2,q,\lambda)\cr
&+g'\;m\;\sum_{\lambda''=0}^2\sum_s\;\zeta_{\lambda''}\;\int d^3k
\;{{\tilde Q}_1(\beta,{\bf n},s_2,s,\lambda'',{\bf k})\over
\sqrt{|{\bf k}|}}\;\phi_{\beta+1}({\bf p},{\bf n}-{\bf
k},s_1,s,q,{\bf k},\lambda,\lambda'')\cr
&-g'\;m\;\sum_{\lambda''=0}^2\sum_s\;\zeta_{\lambda''}\;\int
d^3k\;{{\tilde Q}_2(\beta,{\bf p},s_1,s,\lambda'',{\bf k})\over
\sqrt{|{\bf k}|}} \;\phi_{\beta+1}({\bf p}-{\bf k},{\bf
n},s,s_2,q,{\bf k},\lambda,\lambda'') \cr
&+g'\;m\;\sum_s\;\sum_{k=1}^{\beta}\;{Q_3(\beta,k,{\bf
n},s_2,s)\over \sqrt{|{\bf q}_k|}}\;\phi_{\beta-1}({\bf p},{\bf
n}+{\bf q}_k,s_1,s,{\tilde q_k},{\tilde \lambda_k})\cr
&-g'\;m\;\sum_s\;\sum_{k=1}^{\beta}\; {Q_4(\beta,k,{\bf
p},s_1,s)\over\sqrt{|{\bf q}_k|}}\;\phi_{\beta-1}({\bf p}+{\bf
q}_k,{\bf n},s,s_2,{\tilde q_k},{\tilde\lambda_k})\cr &-{g'^2\over
8}\;\sum_{\lambda',\lambda'''=0}^2\;\zeta_{\lambda'}\;
\zeta_{\lambda'''}\; \sum_{k\ne j}^{\beta}\;\int {d^3k' \over
\sqrt{|{\bf q}_k|\;|{\bf q}_j|\;|{\bf k}'|\; |{\bf q}_k+{\bf
q}_j-{\bf k}'|}}\cr &\times \phi_{\beta}({\bf p},{\bf
n},s_1,s_2,{\bf q}_k+{\bf q}_j- {\bf k}',{\bf k}',{\tilde {{\bf
q}_k}}\; {\tilde {{\bf q}_j}},\lambda''',\lambda', {\tilde
{\lambda_k}}\;{\tilde {\lambda_j}})\;{\tilde
X}(\beta,k,j,\lambda''',\lambda',{\bf k}')=0 &(31)\cr} $$ with the
condition $$ {\bf p}+{\bf n}+\sum_{k=1}^{\beta}{\bf q}_k=0. \eqno
(32) $$ We have defined $$ g'={g_s\over \sqrt{2\;(2\;\pi)^3}} $$
and the definitions for $X$, $I^{\beta}$ and the $Q$'s are given
in APPENDIX A. The factor $X$ is composed of the gluon
polarization vectors, and $Q$ depends on the Dirac spinors as well
as relativistic factors; $I^{\beta}$ is a color sum dependent on
the Gell-Mann matrices. We should point out that the contribution
from the ``cubic'' ${\cal H}_{3g}$ term of equation (20) vanishes
because of the properties of the Gell-Mann matrices and our choice
of variational ansatz (29) which is symmetrized in the Gell-Mann
matrices. Consequently, the only surviving potential terms are of
a color ``Coulombic'' type and a ``quartic'' contribution from the
${\cal H}_{4g}$ term (equation (21)); the nonAbelian effects are
present only for $\beta\ge 2$. If, in our ansatz (29), we remove
the $\{\cdot\cdot\cdot\}$ permutation condition on the Gell-Mann
terms  such that $$ \{t^{z_1}_{l_1\;l_2}\;t^{z_2}_{l_2\;l_3}\cdots
t^{z_{\beta}}_{l_\beta\;{l_{\beta+1}}}\}\rightarrow
t^{z_1}_{l_1\;l_2}\;t^{z_2}_{l_2\;l_3}\cdots
t^{z_{\beta}}_{l_\beta\;{l_{\beta+1}}} $$ then we still retain the
symmetry of $\phi_{\beta}$ under the exchange of any $({\bf
q}_k,\lambda_k)$ pair, but now we obtain a set of equations that
is not as tractable as the present case; that is, it's not obvious
how to decouple them (which we need to do to make the problem
tractable). The ${\cal H}_{3g}$ contribution is now present and
these equations are written explicitly in APPENDIX B for
completeness.
\par\medskip
\centerline {IV. HEAVY-MASS LIMIT}
\par\medskip
The coupled equations of (31) are obviously very difficult to
solve so we will begin with a simplification that will provide an
approximate solution. One can start with the fixed or heavy mass
limit, which will be adequate in dealing with very heavy quark
systems such as $J/\psi$ ($c\overc$) or $\Upsilon$ ($b\overb$)
that have a largely nonrelativistic behavior.
\par
In the lowest order of $\ds {p\over m}$, that is, letting $\ds
\omega_p\rightarrow m+{{\bf p}^2\over 2\;m}$ in the kinetic energy
terms and letting $\omega_p\rightarrow m$ in the potential energy
terms, (31) becomes $$\eqalignno{ &\left[ 2\;m+{{\bf p}^2\over
2\;m}+{{\bf n}^2\over 2\;m}+2\;(m_0-m)+\sum_{k=1}^{\beta}\;|{\bf
q}_k|-E \right]\;I^{\beta}\;\phi_{\beta}({\bf p},{\bf
n},q,\lambda)\cr
&+g'\;\sum_{\lambda''=0}^2\;\zeta_{\lambda''}\;\int d^3k
\;{{\tilde \epsilon}^0_{\lambda''}({\bf k})\over \sqrt{|{\bf
k}|}}\cr &\times \left[J_1^{\beta+1}\;\phi_{\beta+1}({\bf p},{\bf
n}-{\bf k},q, {\bf
k},\lambda,\lambda'')-J_2^{\beta+1}\;\phi_{\beta+1}({\bf p}-{\bf
k},{\bf n},q,{\bf k},\lambda, \lambda'')\right]\cr
&+g'\;\sum_{k=1}^{\beta}\;{\epsilon^0_{\lambda_k}({\bf q}_k)\over
\sqrt{|{\bf q}_k|}}\;\left[J_3^{\beta-1}(z_k)\;\phi_{\beta-1}({\bf
p},{\bf n}+{\bf q}_k, {\tilde
q_k},{\tilde\lambda_k})-J_4^{\beta-1}(z_k)\;\phi_{\beta-1}({\bf
p}+{\bf q}_k,{\bf n},{\tilde q_k},{\tilde\lambda_k})\right]\cr
&-{g'^2\over
8}\;\sum_{\lambda',\lambda'''=0}^2\;\zeta_{\lambda'}\;
\zeta_{\lambda'''}\; \sum_{k\ne j}^{\beta}\;\int {d^3k' \over
\sqrt{|{\bf q}_k|\;|{\bf q}_j|\;|{\bf k}'|\; |{\bf q}_k+{\bf
q}_j-{\bf k}'|}}\cr &\times \phi_{\beta}({\bf p},{\bf n},{\bf
q}_k+{\bf q}_j- {\bf k}',{\bf k}',{\tilde {{\bf q}_k}}\; {\tilde
{{\bf q}_j}},\lambda''',\lambda', {\tilde {\lambda_k}}\;{\tilde
{\lambda_j}})\;{\tilde X}(\beta,k,j,\lambda''',\lambda',{\bf
k}')=0 & (33)\cr} $$ where $$\eqalignno{ &{\tilde
\epsilon}^0_{\lambda''}({\bf k})= \epsilon^0_0({\bf
k})+\epsilon^0_3({\bf k})&(34)\cr} $$ for $\lambda''=0$, and $$
\eqalignno{ &{\tilde \epsilon}^0_{\lambda''}({\bf k})=
\epsilon^0_{\lambda''}({\bf k})&(35)\cr} $$ for $\lambda''=1,2$.
\par
In the QED fixed-mass case, the kinetic energy terms are neglected
and all the $I$'s and $J$'s in (33) collapse to $1$, thus allowing
the infinite sequence of coupled equations to have an exact
solution [13]: an ansatz, $\phi_{\beta}$, may be found (along with
a specific mass renormalization condition) that decouples the
equations. However, in the QCD case the Gell-Mann matrices hamper
such an exact solution, but we will follow the spirit of that
approach in at least getting an approximate solution. Thus we
choose an ansatz $$\eqalignno{ &I^{\beta}\;\phi_{\beta}({\bf
p},{\bf n},q,\lambda)= -g'\;{\epsilon^0_{\lambda_{\beta}}({\bf
q}_{\beta})\over \sqrt{|{\bf q}_{\beta}|^3}}\cr & \times
\left[J_3^{\beta-1}(z_{\beta})\;\phi_{\beta-1}({ \bf p},{\bf
n}+{\bf q}_{\beta},{\tilde
q_{\beta}},{\tilde\lambda_{\beta}})-J_4^{\beta-1}(z_{\beta})\;
\phi_{\beta-1}({\bf p}+{\bf q}_{\beta},{\bf n},{\tilde
q_{\beta}},{\tilde\lambda_{\beta}})\right],&(36)\cr} $$ substitute
it in (33), and impose a mass-renormalization condition
($\beta$-dependent) $$ 2\;(m_0-m)=-{\alpha\over
4\pi^2}\;\Biggl\{{J_1^{\beta+1}\;J_3^{\beta}(z_{\beta+1})\over
I^{\beta}\;I^{\beta+1}}+
{J_2^{\beta+1}\;J_4^{\beta}(z_{\beta+1})\over
I^{\beta}\;I^{\beta+1}}\Biggr\} \int {d^3k\over|{\bf k}|^2}
\eqno(37) $$ This uncouples the chain of equations (33) and
results in a sequence of bound-state equations $$\eqalignno{
&\left[ 2\;m+{{\bf p}^2\over 2\;m}+{{\bf n}^2\over
2\;m}-E\right]\;\phi_{\beta}({\bf p},{\bf n},q,\lambda)\cr
&-{\alpha\over 4\pi^2}   \;\Biggl\{
{J_1^{\beta+1}\;J_4^{\beta}(z_{\beta+1})\over
I^{\beta}\;I^{\beta+1}}+
{J_2^{\beta+1}\;J_3^{\beta}(z_{\beta+1})\over
I^{\beta}\;I^{\beta+1}} \Biggr\} \int {d^3k\over |{\bf
k}|^2}\;\phi_{\beta}({\bf p}+{\bf k},{\bf n}-{\bf k},q,\lambda)\cr
&-{\alpha\over
32\;\pi^2\;I^{\beta}}\;\sum_{\lambda',\lambda'''=0}^2\;
\zeta_{\lambda'}\;\zeta_{\lambda'''}\; \sum_{k\ne j}^{\beta}\;\int
{d^3k' \over \sqrt{|{\bf q}_k|\;|{\bf q}_j|\;|{\bf k}'|\; |{\bf
q}_k+{\bf q}_j-{\bf k}'|}}\cr &\times \phi_{\beta}({\bf p},{\bf
n},{\bf q}_k+{\bf q}_j- {\bf k}',{\bf k}',{\tilde {{\bf q}_k}}\;
{\tilde {{\bf q}_j}},\lambda''',\lambda', {\tilde
{\lambda_k}}\;{\tilde {\lambda_j}})\;{\tilde
X}(\beta,k,j,\lambda''',\lambda',{\bf k}')=0 &(38)\cr} $$ with the
correct rest-plus-kinetic energy for a two-particle system. It is
understood that the divergent integral in (37) is controlled by a
suitable regulator (cut-off) which does not appear, subsequently,
in (38).
\par
Equation (38), for any given $\beta$, is a momentum space
Schr\"odinger-like equation for the stationary states of the heavy
quark-antiquark system; there is an obvious color ``Coulomb'' term
present plus a ``quartic'' contribution from the ${\cal H}_{4g}$
term. On the surface, this ``quartic'' term also seems to share a
Coulomb-type ($\ds {1\over |{\bf k}|^2}$) behavior. An explicit
$\ds {1\over |{\bf k}|^3}$ or $\ds {1\over |{\bf k}|^4}$ term in
the ``quartic'' part does not appear, a term that would signal a
confining-type potential.
\par
Note that we have chosen a specific representation [12]:
$$\eqalignno{ &\epsilon^{\mu}_0({\bf k})=\eta^{\mu}=(1,0,0,0)\cr
&\epsilon^{\mu}_r({\bf k})=(0,\fakebold{$\epsilon$}_r({\bf
k})),\;\;\;r=1,2,3\cr
&\fakebold{$\epsilon$}_1\cdot\fakebold{$\epsilon$}_2=0,\;\;\;
\fakebold{$\epsilon$}_3({\bf k})={{\bf k}\over|{\bf k}|},\cr &{\bf
k}\cdot\fakebold{$\epsilon$}_r({\bf k})=0,\;\;r=1,2 &(39)\cr} $$
with $$ \alpha={g_s^2\over 4\;\pi} $$ to simplify the color
``Coulomb'' term. We'll work out the terms for $X$ in the
``quartic'' term shortly.
\par
The sequence of equations (38) represents different approximations
for the $q\overq$ system with various numbers, $\beta$, of
``spectator'' gluons. Note that nonAbelian effects begin only with
$\beta\ge 2$. Explicitly, for $\beta=0,1,2$, our
mass-renormalization conditions, bound-state equations and momenta
constraints are $$ 2\;(m_0-m)=-{\alpha\over 2\pi^2}\;{4\over 3}
\int {d^3k\over|{\bf k}|^2} \eqno(40a) $$ $$\eqalignno{ &\left[
2\;m+{{\bf p}^2\over m} -E\right]\;\phi_{0}({\bf p},{\bf
n})-{\alpha\over 2\pi^2}   \;{ 4\over 3} \int {d^3k\over |{\bf
k}|^2}\;\phi_{0}({\bf p}+{\bf k},{\bf n}-{\bf k})=0\cr &
&(40b)\cr} $$ $$ {\bf p}+{\bf n}=0 \eqno (40c) $$ for $\beta=0$,
$$ 2\;(m_0-m)=-{\alpha\over 2\pi^2}\;{7\over 12}\;\int
{d^3k\over|{\bf k}|^2} \eqno(41a) $$ $$\eqalignno{ &\left[
2\;m+{{\bf p}^2\over 2\;m}+{{\bf n}^2\over
2\;m}-E\right]\;\phi_{1}({\bf p},{\bf n},{\bf
q}_1,\lambda_1)-{\alpha\over 2\pi^2} \;{7\over 12}\;\int
{d^3k\over |{\bf k}|^2}\;\phi_{1}({\bf p}+{\bf k},{\bf n}-{\bf
k},{\bf q}_1,\lambda_1)=0\cr & &(41b)\cr} $$ $$ {\bf p}+{\bf
n}+{\bf q}_1=0 \eqno(41c) $$ for $\beta=1$, and $$
2\;(m_0-m)=-{\alpha\over 2\pi^2}\;{10\over 21} \int
{d^3k\over|{\bf k}|^2} \eqno(42a) $$ $$\eqalignno{ &\left[
2\;m+{{\bf p}^2\over 2\;m}+{{\bf n}^2\over
2\;m}-E\right]\;\phi_{2}({\bf p},{\bf n},{\bf q}_1,{\bf
q}_2,\lambda_1, \lambda_2)\cr &-{\alpha\over 2\pi^2}\;{10\over 21}
\int {d^3k\over |{\bf k}|^2}\;\phi_{2}({\bf p}+{\bf k},{\bf
n}-{\bf k},{\bf q}_1,{\bf q}_2,\lambda_1,\lambda_2)\cr
&-{3\;\alpha\over 32\;\pi^2\;56}
\;\sum_{\lambda',\lambda'''=0}^2\;\zeta_{\lambda'}\;
\zeta_{\lambda'''}\; \int {d^3k' \over \sqrt{|{\bf q}_1|\;|{\bf
q}_2|\;|{\bf k}'|\; |{\bf q}_1+{\bf q}_2-{\bf k}'|}}\cr &\times
\phi_{2}({\bf p},{\bf n},{\bf q}_1+{\bf q}_2- {\bf k}',{\bf
k}',\lambda''',\lambda')\Biggl[\;{\tilde
X}(2,1,2,\lambda''',\lambda',{\bf k}')+{\tilde
X}(2,2,1,\lambda''',\lambda',{\bf k}')\Biggr]\cr &=\left[
2\;m+{{\bf p}^2\over 2\;m}+{{\bf n}^2\over
2\;m}-E\right]\;\phi_{2}({\bf p},{\bf n},{\bf q}_1,{\bf
q}_2,\lambda_1, \lambda_2)\cr &-{\alpha\over 2\pi^2}\;{10\over 21}
\int {d^3k\over |{\bf k}|^2}\;\phi_{2}({\bf p}+{\bf k},{\bf
n}-{\bf k},{\bf q}_1,{\bf q}_2,\lambda_1,\lambda_2)\cr
&-{6\;\alpha\over 32\;\pi^2\;56}
\;\sum_{\lambda',\lambda'''=0}^2\;\zeta_{\lambda'}\;
\zeta_{\lambda'''}\; \int {d^3k' \over \sqrt{|{\bf q}_1|\;|{\bf
q}_2|\;|{\bf k}'|\; |{\bf q}_1+{\bf q}_2-{\bf k}'|}}\cr &\times
\phi_{2}({\bf p},{\bf n},{\bf q}_1+{\bf q}_2- {\bf k}',{\bf
k}',\lambda''',\lambda'){\tilde X}(2,1,2,\lambda''',\lambda',{\bf
k}')=0, &(42b)\cr} $$ $$ {\bf p}+{\bf n}+{\bf q}_1+{\bf q}_2=0
\eqno (42c) $$ for $\beta=2$. There are no gluon kinetic energy
terms in (41b) and (42b) because of the ansatz (36).
\par
All three equations have a one-gluon exchange ``Coulomb'' term,
modified by color factors stemming from the sums explicitly
written in APPENDIX A; that is, we have the coefficients ${4\over
3}$, ${7\over 12}$ and ${10\over 21}$ for $\beta=0,1,2$
respectively. The $\beta=0$ equation has only the virtual
one-gluon exchange and its familiar ${4\over 3}$ factor, while the
$\beta=1,2$ bound state equations have spectator gluons present.
However it is only for $\beta= 2$ (more generally $\beta\ge 2$)
that there  appears a nonAbelian contribution to the interquark
potential energy, and so we expect that the $\beta=2$ equation is
the more realistic representation of quarkonium from among the
three ($\beta=0,1,2$) approximations. It is evident, though, that
the equations become increasingly more complicated, and so more
difficult to solve, as $\beta$ increases.
\par\medskip
\centerline {V. APPROXIMATE SOLUTIONS}
\par\medskip
\centerline {A. Suppression of NonAbelian terms}
\par\medskip
If we ignore the nonAbelian terms in equation (38), we have
$$\eqalignno{ &\left[ 2\;m+{{\bf p}^2\over 2\;m}+{{\bf n}^2\over
2\;m}-E\right]\;\phi_{\beta}({\bf p},{\bf
n},q,\lambda)-{\alpha\over 4\pi^2}   \;\gamma(\beta)\; \int
{d^3k\over |{\bf k}|^2}\;\phi_{\beta}({\bf p}+{\bf k},{\bf n}-{\bf
k},q,\lambda)=0 &(43)\cr} $$ with $$ {\bf p}+{\bf
n}+\sum_{k=1}^{\beta}{\bf q}_k=0, \eqno (44) $$ and where (see
APPENDIX A) $$ \eqalignno{ \gamma(\beta)&=\Biggl\{
{J_1^{\beta+1}\;J_4^{\beta}(z_{\beta+1})\over
I^{\beta}\;I^{\beta+1}}+
{J_2^{\beta+1}\;J_3^{\beta}(z_{\beta+1})\over
I^{\beta}\;I^{\beta+1}}
\Biggr\}=\Biggl\{{J_1^{\beta+1}\;J_3^{\beta}(z_{\beta+1})\over
I^{\beta}\;I^{\beta+1}}+
{J_2^{\beta+1}\;J_4^{\beta}(z_{\beta+1})\over
I^{\beta}\;I^{\beta+1}}\Biggr\}\cr &=2\;{I^{\beta+1}\over
(\beta+1)^3\;I^{\beta}}. &(45)\cr} $$ Furthermore, unlike similar
equations in QED, we note that $\gamma(\beta)$ is not a constant
but $\beta$-dependent; that is, it is determined by the number of
``spectator'' gluons. The first few terms are $$\eqalignno{
\gamma(0)&=2\cdot{4\over 3}=2\;(1.3333)\cr
\gamma(1)&=2\cdot{7\over 12}=2\;(0.5833)\cr
\gamma(2)&=2\cdot{10\over 21}=2\;(0.4762)\cr
\gamma(3)&=2\cdot{47\over 96}=2\;(0.4896)\cr
\gamma(4)&=2\cdot{62\over 141}=2\;(0.4397)\cr
\gamma(5)&=2\cdot{1621\over 3720}=2\;(0.4357) &(45a)\cr} $$ which
reveal a screening effect caused by the spectator gluons, with
$\gamma(\beta)$ apparently tending to a limiting value. In the QED
case, we would have $\gamma(\beta)=2$ for all $\beta$.
\par
Equation (43) has exact solutions for each $\beta$, as could be
found easily in coordinate space. For example, the $\beta=0$ case
of equation (43) is presented in equation (40b) and may be
rewritten as $$\eqalignno{ &\left[ {{\bf p}^2\over m}
-\epsilon\right]\;\phi_{0}({\bf p})-{\alpha\over 4\pi^2}
\;\gamma(0) \int {d^3k\over |{\bf k}|^2}\;\phi_{0}({\bf p}+{\bf
k})=0 &(46)\cr} $$ where $\ds \phi_0({\bf p})$ represents $\ds
\phi_0({\bf p},{\bf n})$ (recall that ${\bf p}+{\bf n}=0$ for
$\beta=0$). In coordinate space, equation (46) transforms to the
familiar $$ \Biggl(-{{\bigtriangledown^2_{r}}\over 2\;\mu}
-{\alpha\;\gamma(0)\over 2\;|{\bf r}|}\Biggr)\;\phi_0({\bf r})
=\epsilon\;\phi_0({\bf r}) \eqno (47) $$ where $$ \mu={m\over 2}
$$ and $$ \epsilon=E-2\;m. $$ Equation (47) has the standard
(modified) solutions $$
\epsilon=E-2\;m=-{\mu\;\alpha^2\;\gamma^2(0)\over 8\;n^2}. \eqno
(48) $$ The $\ds \phi_0({\bf r})$ of (47) correspond to the
momentum space wave functions $$ \phi_0({\bf p})\propto {p^l\over
(p^2+a^2)^{2+l}}\;C^{l+1}_{n-l-1}({p^2-a^2\over p^2+a^2}) \eqno
(49) $$ with $$ a={\mu\;\alpha\;\gamma(0)\over 2\;n} \eqno (50) $$
and the $C^{l+1}_{n-l-1}$'s are the Gegenbauer functions.
\par
Now, for $\beta=1$ (which corresponds to a term with the
$q\;\overq$ pair plus an explicit gluon; i.e., an $H_2^{+}$-like
configuration), equation (43) yields equation (41b), which we
rewrite as $$\left[ {{\bf p}^2\over 2\;m}+{{\bf n}^2\over
2\;m}-\epsilon\right]\;\phi_{1}({\bf p},{\bf
n},\lambda_1)-{\alpha\over 4\pi^2} \;{\gamma(1)}\;\int {d^3k\over
|{\bf k}|^2}\;\phi_{1}({\bf p}+{\bf k},{\bf n}-{\bf
k},\lambda_1)=0 \eqno(51) $$ with $\ds \phi_{1}({\bf p},{\bf
n},\lambda_1)\equiv \phi_{1}({\bf p},{\bf n},{\bf q_1},\lambda_1)$
(recall that ${\bf p}+{\bf n}+{\bf q}_1=0$ for this case). If we
assume that $$\phi_{1}({\bf p},{\bf n},\lambda_1)=\phi_{1}({\bf
p},{\bf n})\;Z(\lambda_1) $$ then the $Z$ factors out and in
coordinate space (51) is $$\Biggl({1\over
2\;m}\;(-\bigtriangledown^2_{r_1}-\bigtriangledown^2_{r_2})
-{\alpha\;\gamma(1)\over 2\;|{\bf r}_1-{\bf
r}_2|}\;\Biggr)\;\phi_1({\bf r}_1,{\bf r}_2)=\epsilon\;\phi_1({\bf
r}_1,{\bf r}_2) \eqno (52) $$ If we transform to new coordinates
$$ {\bf r}={\bf r}_1-{\bf r}_2, \eqno (53) $$ $$ {\bf R}={{\bf
r}_1+{\bf r}_2\over 2} \eqno (54) $$ and let $$ \phi_1({\bf
r}_1,{\bf r}_2)=\phi_1({\bf r})\;\phi_1({\bf R}) \eqno (55) $$
equation (52) reduces to the form (47), with a solution $$
\epsilon=E-2\;m=-{\mu\;\alpha^2\;\gamma^2(1)\over 8\;n^2}.  \eqno
(56) $$ Here, $\ds \phi_1({\bf r})$ is the typical hydrogenic
wavefunction, and $\ds \phi_1({\bf R})$ is a plane wave solution.
In momentum space, $$ \phi_1({\bf r})\;\phi_1({\bf R})\rightarrow
\phi_1({\bf p_r})\;\delta({\bf P_R}) \eqno (57) $$ where $\ds
\phi_1({\bf p_r})$ is the typical momentum space hydrogenic
wavefunction (see (49)). Noting the transformations (53) and (54),
the analogous procedure in momentum space is to let $$ {\bf
p_r}={{\bf p}-{\bf n}\over 2} \eqno (58) $$ and $$ {\bf P_R}={{\bf
p}+{\bf n}}. \eqno (59) $$ This will be of use in the next
section.
\par
Lastly, for $\beta=2$ (a system that is composed of a $q\;\overq$
pair plus two explicit gluons; i.e., an $H_2$-molecule-like
configuration), equation (43) yields (equation (42b) without the
nonAbelian terms) $$ \left[ {{\bf p}^2\over 2\;m}+{{\bf n}^2\over
2\;m}-\epsilon\right]\;\phi_{2}({\bf p},{\bf n},{\bf
q}_1)-{\alpha\over 4\pi^2}\;{\gamma(2)} \int {d^3k\over |{\bf
k}|^2}\;\phi_{2}({\bf p}+{\bf k},{\bf n}-{\bf k},{\bf q}_1)=0
\eqno (60) $$ with $\phi_{2}({\bf p},{\bf n},{\bf q}_1)\equiv
\phi_{2}({\bf p},{\bf n},{\bf q}_1,{\bf q}_2)$ (recall ${\bf
p}+{\bf n}+{\bf q}_1+{\bf q}_2=0$) and we have factored out the
$\lambda$ terms as in the last case. In coordinate space, equation
(60) is $$ \Biggl({1\over
2\;m}\;(-\bigtriangledown^2_{r_1}-\bigtriangledown^2_{r_2})
-{\alpha\;\gamma(2)\over 2\;|{\bf r}_1-{\bf
r}_2|}\;\Biggr)\;\phi_2({\bf r}_1,{\bf r}_2,{\bf
r}_3)=\epsilon\;\phi_2({\bf r}_1,{\bf r}_2,{\bf r}_3). \eqno (61)
$$ Note that ${\bf r}_3$ appears as a parameter in equation (61);
it has no effect on $\epsilon$. As for the $\beta=1$ case, we can
use the transformations (53) and (54) to get $$
\epsilon=E-2\;m=-{\mu\;\alpha^2\;\gamma^2(2)\over 8\;n^2}, \eqno
(62) $$ where the wave functions have the same structure as
$\beta=1$. We expect the same result to hold for $\beta
>2$, with arbitrary functions present. That is, in general, we
have $$ \epsilon=E-2\;m=-{\mu\;\alpha^2\;\gamma^2(\beta)\over
8\;n^2} \eqno (63) $$ (with an `inverse Bohr radius' $\ds
a={\mu\;\alpha\;\gamma(\beta)\over 2\;n}$ used in the respective
momentum space wave functions).
\par
Let us apply this modified Balmer formula to the heavy quark
mesons such as charmonium and bottomonium to obtain a prediction
for the low-lying bound states. Using the experimental values for
the lowest-lying $1S$ and $2S$ states of $c\overc$ (i.e.,
$J/\psi(1S)$, $J/\psi(2S)$; [11]), we can fix two of our
parameters; that is, we can set our mass for the charm quark
($m_c$) as well as the coupling constant $\alpha$ particular to
this system. Once this is done, we use our expressions for $E$
(equations (48),(56), and (62)) to get the predicted values for
the $3S$, $4S$, $5S$ and $6S$ states. We repeat the process for
the $b\overb$ mesons, using $\Upsilon(1S)$, $\Upsilon(2S)$, and
present the results in Tables I and II for charmonium and
bottomonium respectively.
\par
In this approximation suppressing the nonAbelian terms, we note
that the predicted values for charmonium and bottomonium are
independent of $\beta$, the number of ``spectator'' gluons
present. The mass values are constant, as can be easily seen from
equation (61) for $n=1,2$, as is the value
$\alpha^2\;\gamma^2(\beta)$; only $\alpha$ changes with ${\beta}$.
As expected, the predicted masses are not particularly close to
the observed ones, though the values for $b\overb$ are better than
for charmonium, since this is a heavier, more nonrelativistic
system for which the mass spectrum is influenced more by the
Coulomb potential. One would expect better results still for
$t\overt$, were the data available.
\par\medskip
\centerline {B. NonAbelian terms present: $\beta=2$ case}
\par\medskip
We now turn to the more realistic approximation in which we have
an explicit manifestation of the nonAbelian nature of the
interaction. Recalling our transformations (58) and (59), our
equation (42b), for $\beta=2$, may be rewritten in the new
coordinates as $$\eqalignno{ &\left[ {{\bf p_r}^2\over
2\;\mu}+{{\bf P_R}^2\over 8\;\mu}-\epsilon\right]\;\phi_{2}({\bf
p_r},{\bf P_R},{\bf q}_1,{\bf q}_2,\lambda_1,\lambda_2)\cr
&-{\alpha\over 4\pi^2}\;\gamma(2)\; \int {d^3k\over |{\bf
k}|^2}\;\phi_{2}({\bf p_r}-{\bf k},{\bf P_R},{\bf q}_1,{\bf
q}_2,\lambda_1,\lambda_2)\cr &-{6\;\alpha\over 32\;\pi^2\;56}
\;\sum_{\lambda',\lambda'''=0}^2\;\zeta_{\lambda'}\;
\zeta_{\lambda'''}\; \int {d^3k' \over \sqrt{|{\bf q}_1|\;|{\bf
q}_2|\;|{\bf k}'|\; |{\bf q}_1+{\bf q}_2-{\bf k}'|}}\cr &\times
\phi_{2}({\bf p_r},{\bf P_R},{\bf q}_1+{\bf q}_2- {\bf k}',{\bf
k'},\lambda''',\lambda'){\tilde X}(2,1,2,\lambda''',\lambda',{\bf
k}')=0&(64)\cr} $$
\par
If we make the transformation $$ \sqrt{|{\bf q}_1}|\;\sqrt{|{\bf
q}_2|}\;\phi_{2}({\bf p_r},{\bf P_R},{\bf q}_1,{\bf
q}_2,\lambda_1,\lambda_2)= \Phi({\bf p_r},{\bf P_R},{\bf q}_1,{\bf
q}_2)\;Z(\lambda_1,\lambda_2) \eqno (65) $$ then equation (64)
becomes $$\eqalignno{ &\left[  {{\bf p_r}^2\over 2\;\mu}+{{\bf
P_R}^2\over 8\;\mu}-\epsilon\right]\;\Phi({\bf p_r},{\bf P_R},{\bf
q}_1,{\bf q}_2)\;Z(\lambda_1, \lambda_2)\cr &-{\alpha\over
4\pi^2}\;\gamma(2)\; \int {d^3k\over |{\bf k}|^2}\;\Phi({\bf
p_r}-{\bf k},{\bf P_R},{\bf q}_1,{\bf
q}_2)\;Z(\lambda_1,\lambda_2)\cr &-{6\;\alpha\over 32\;\pi^2\;56}
\;\sum_{\lambda',\lambda'''=0}^2\;\zeta_{\lambda'}\;
\zeta_{\lambda'''}\; \int {d^3k' \over |{\bf k}'|\;|{\bf q}_1+{\bf
q}_2-{\bf k}'|}\cr &\times \Phi({\bf p_r},{\bf P_R},{\bf q}_1+{\bf
q}_2- {\bf k}',{\bf k}')\;Z(\lambda''',\lambda'){\tilde
X}(2,1,2,\lambda''',\lambda',{\bf k}')=0& (66)\cr} $$ For an
approximate solution, let's multiply (66) by $\ds \Phi({\bf
p_r},{\bf P_R},{\bf q}_1,{\bf q}_2)$ and integrate over ${\bf
p_r},{\bf P_R},{\bf q_1}$ such that $$\eqalignno{
&\int\;d^3p_r\;d^3P_R\;d^3q_1\;\left[  {{\bf p_r}^2\over
2\;\mu}+{{\bf P_R}^2\over 8\;\mu}-\epsilon\right]\;\Phi^{2}({\bf
p_r},{\bf P_R},{\bf q}_1,{\bf q}_2)\;Z(\lambda_1, \lambda_2)\cr
&-{\alpha\over 4\pi^2}\;\gamma(2)\; \int {d^3k\over |{\bf
k}|^2}\;d^3p_r\;d^3P_R\;d^3q_1\;\Phi({\bf p_r},{\bf P_R},{\bf
q}_1,{\bf q}_2)\;\Phi({\bf p_r}-{\bf k},{\bf P_R},{\bf q}_1,{\bf
q}_2)\;Z(\lambda_1,\lambda_2)\cr &-{6\;\alpha\over 32\;\pi^2\;56}
\;\sum_{\lambda',\lambda'''=0}^2\;\zeta_{\lambda'}\;
\zeta_{\lambda'''}\; \int {d^3k'\;d^3p_r\;d^3P_R\;d^3q_1 \over
|{\bf k}'|\;|{\bf q}_1+{\bf q}_2-{\bf k}'|}\cr &\times \Phi({\bf
p_r},{\bf P_R},{\bf q}_1,{\bf q}_2)\;\Phi({\bf p_r},{\bf P_R},{\bf
q}_1+{\bf q}_2- {\bf k}',{\bf k}')\;Z(\lambda''',\lambda'){\tilde
X}(2,1,2,\lambda''',\lambda',{\bf k}')=0& (67)\cr} $$ As a trial
function, we will use the nonAbelian solutions of the last section
$$ \Phi({\bf p_r}\,{\bf P_R},{\bf q}_1,{\bf q}_2)=\phi_c({\bf
p_r})\;\delta({\bf P_R})\;\psi({\bf q}_1) \eqno (68) $$ where $\ds
\phi_c({\bf p_r})$ is the classic hydrogenic wave function and
$\ds \psi({\bf q}_1)$ is an arbitrary function which we choose to
regulate the integral in momentum space. We can treat $a$ in the
wave function as a variational parameter to optimize our results
(see (49),(50)). Integrating over ${\bf P_R}$ leads us to
$$\eqalignno{ &\int\;d^3p_r\;\left[  {{\bf p_r}^2\over
2\;\mu}-\epsilon\right]\;\phi^{2}_c({\bf
p_r})\;Z(\lambda_1,\lambda_2)\;\int\;d^3q_1\;\psi^2({\bf q}_1)\cr
&-{\alpha\over 4\pi^2}\;\gamma(2)\; \int {d^3k\over |{\bf
k}|^2}\;d^3p_r\;\phi_c({\bf p_r})\;\phi_c({\bf p_r}-{\bf
k})\;Z(\lambda_1,\lambda_2)\;\int d^3q_1\;\psi^2({\bf q}_1)\cr
&-{6\;\alpha\over
32\;\pi^2\;56}\;\sum_{\lambda',\lambda'''=0}^2\;\zeta_{\lambda'}
\;\zeta_{\lambda'''}\;\int\; d^3p_r\;\phi^{2}_c({\bf
p_r})\;Z(\lambda''',\lambda')\;\int\;{d^3k'\;\psi(-{\bf k}')\over
|{\bf k}'|^2}\cr &\times \int\;d^3 q_1\;\psi({\bf q}_1)\;{\tilde
X}(2,1,2,\lambda''',\lambda',{\bf k}')=0& (69)\cr} $$ with $$ {\bf
q}_1+{\bf q}_2+{\bf P_R}={\bf q}_1+{\bf q}_2=0; $$ i.e., $$ {\bf
q}_2=-{\bf q}_1 $$ which simplifies the polarization vector terms
in $X$ considerably (see APPENDIX A, equation (A8)). Summing over
$\lambda_1,\lambda_2$ in (69), we obtain $$\eqalignno{
&\int\;d^3p_r\;\left[  {{\bf p^2_r}\over
2\;\mu}-\epsilon\right]\;\phi^{2}_c({\bf
p_r})\;\int\;d^3q_1\;\psi^2({\bf q}_1)\cr &-{\alpha\over
4\pi^2}\;\gamma(2)\; \int {d^3k\over |{\bf
k}|^2}\;d^3p_r\;\phi_c({\bf p_r})\;\phi_c({\bf p_r}-{\bf k})\;\int
d^3q_1\;\psi^2({\bf q}_1)\cr &-{6\;\alpha\over
32\;\pi^2\;56}\;\sum_{\lambda',\lambda'''=0}^2\;\zeta_{\lambda_1}
\;\zeta_{\lambda_2}\;\int\; d^3p_r\;\phi^{2}_c({\bf
p_r})\;\int\;{d^3k'\;\psi(-{\bf k}')\over |{\bf k}'|^2}\cr &\times
\int\;d^3 q_1\;\psi({\bf q}_1)\;{\tilde
X}(2,1,2,\lambda''',\lambda',{\bf
k}')_{\lambda'''\leftrightarrow\lambda_1,\;\lambda'\leftrightarrow\lambda_2}=0
& (70)\cr} $$ Summing again to remove the $\lambda_1,\lambda_2$
terms still present in the ``quartic'' term of (70), that is,
working out $$
\sum_{\lambda',\lambda'''=0}^2\;\sum_{\lambda_1,\lambda_2=0}^2\;
\zeta_{\lambda_1}\;\zeta_{\lambda_2}\;{\tilde
X}(2,1,2,\lambda''',\lambda',{\bf
k}')_{\lambda'''\leftrightarrow\lambda_1,\;\lambda'\leftrightarrow\lambda_2}\eqno
(71) $$ using the convention of [14] $$
\fakebold{$\epsilon$}_1(-{\bf k})=-\fakebold{$\epsilon$}_1({\bf
k}),\;\;\fakebold{$\epsilon$}_2(-{\bf
k})=\fakebold{$\epsilon$}_2({\bf k} ) \eqno(72) $$ and recalling
(39), we can simplify (70) to $$\eqalignno{ &\int\;d^3p_r\;\left[
{{\bf p^2_r}\over 2\;\mu}-\epsilon\right]\;\phi^{2}_c({\bf
p_r})\;\int\;d^3q_1\;\psi^2({\bf q}_1)\cr &-{\alpha\over
4\pi^2}\;\gamma(2)\; \int {d^3k\over |{\bf
k}|^2}\;d^3p_r\;\phi_c({\bf p_r})\;\phi_c({\bf p_r}-{\bf k})\;\int
d^3q_1\;\psi^2({\bf q}_1)\cr &-{12\cdot 36\;\alpha\over
32\;\pi^2\;56}\;\int\; d^3p_r\;\phi^{2}_c({\bf
p_r})\;\int\;{d^3k'\;\psi(-{\bf k}')\over |{\bf k}'|^2}\;\int\;d^3
q_1\;\psi({\bf q}_1)\cr &\times \Biggl\{ \Biggl(
\fakebold{$\epsilon$}_{1}({\bf
k}')\cdot\fakebold{$\epsilon$}_{1}({\bf q}_1)+
\fakebold{$\epsilon$}_{2}({\bf
k}')\cdot\fakebold{$\epsilon$}_{2}({\bf q}_1)-
\fakebold{$\epsilon$}_{1}({\bf q}_1)\cdot{{\bf k}'\over |{\bf
k}'|} \Biggr)^2\cr &-\Biggl( \fakebold{$\epsilon$}_{1}({\bf
k}')\cdot\fakebold{$\epsilon$}_{2}({\bf q}_1)+
\fakebold{$\epsilon$}_{2}({\bf
k}')\cdot\fakebold{$\epsilon$}_{1}({\bf q}_1)-
\fakebold{$\epsilon$}_{2}({\bf q}_1)\cdot{{\bf k}'\over |{\bf
k}'|}\Biggr)^2\Biggr\}=0& (73)\cr} $$ or, $$\eqalignno{
&\epsilon=\Biggl\{\int\;d^3p_r\;{{\bf p^2_r}\over
2\;\mu}\;\phi^{2}_c({\bf p_r})\;-{\alpha\over 4\pi^2}\;\gamma(2)\;
\int {d^3k\over |{\bf k}|^2}\;d^3p_r\;\phi_c({\bf
p_r})\;\phi_c({\bf p_r}-{\bf k})\Biggr\}\bigg/ \;
\int\;d^3p_r\;\phi^{2}_c({\bf p_r})\cr &-{18\;\alpha\over 7}\;
\int\;dk'\;\psi(-{\bf k}')\;\int\;dq_1\;q_1^2\;\psi({\bf
q}_1)\bigg/\; \int\;d^3q_1\;\psi^2({\bf q}_1) &(74)\cr} $$
\par
We will use trial functions of the scaled hydrogenic-type
(equation (49)); for example, for the ground state we have
$$\phi_c({\bf p_r})\propto {1\over (p_r^2+a^2)^2}, $$ where $a$ is
an arbitrary scale parameter. For the trial function $\psi({\bf
q}_1)$, we choose a form $$\psi(a,{\bf q}_1)={1\over
(q_1^2+a^2)^2},\;\;\;\;e^{-{q_1\over a}},$$ using the same scale
$a$ as for $\phi_c({\bf p_r})$. In the second part of (74),
changing the variable by letting $\ds q'_1={q_1\over a}$ allows us
to pull out the $a$. That is, $$ \epsilon={a^2\over
m}-{a\;\alpha\;\gamma(2)\over 2\;n}-{9\;a\;\alpha\over 14\;\pi}\;Y
\eqno (75) $$ where we define $$ Y={\int\;dq_1'\;\psi(-{\bf
q}_1')\;\int\;dq_1'\;{q'_1}^2\psi({\bf q}_1')\over
\int\;dq_1'\;{q'_1}^2\psi^2({\bf q}'_1)}. \eqno (76) $$ If we
optimize (75) with respect to $a$ we obtain the optimum value$$
a_n={m\;\alpha\over 2}\;\Biggl( {10\over 21\;n}+{9\;Y\over
14\;\pi} \Biggr) \eqno (77) $$ and so the corresponding$$
\epsilon_n=-{m\;\alpha^2\over 4}\;\Biggl( {10\over
21\;n}+{9\;Y\over 14\;\pi}  \Biggr)^2. \eqno (78) $$ Note that the
binding energy is non-zero as $n\rightarrow \infty$, quite unlike
the Abelian (QED-like) case.
\par
We try a function $$ \psi({\bf q}'_1)={1\over (q_1'^2+1)^N} \eqno
(79) $$ in (78) and use the $1S$ and $2S$ levels to fix $\alpha$
and $m$. Thus we get the predicted values for the $3S$, $4S$, $5S$
and $6S$ states for $c\overc$ which are listed in Table III and
for $b\overb$ in Table IV. Alternatively, trying a function of
type $$ \psi({\bf q}_1')=e^{-q'^N} \eqno (80) $$ produces the
results in Table V and Table VI, for $c\overc$ and $b\overb$
respectively. These results are for the $\beta=2$ case only and we
find that the inclusion of the nonAbelian terms improves the
agreement with experiment substantially over the results of Tables
I and II which were derived using only the modified Balmer formula
(nonAbelian terms suppressed). Once again, the bottomonium values
are better than the charmonium ones. The trial function (80) seems
somewhat better than (79), but the difference is not large,
suggesting that the approximate variational solutions are
reasonably accurate. Our results vary with our choice of parameter
$N$, with $N=1.75$ giving the best results (note that for our
trial function (79), we must have $N > 1.5$ to insure convergence
of our integrals in (76); pushing down to this limit improves the
results slightly. For (80), we need $N>0$).
\par\medskip
\centerline {C. NonAbelian terms present: $\beta=3,4,5$ cases}
\par\medskip
For the $\beta=3,4$, and $5$ cases, we follow the same procedure
as in the previous section. Our equations become more complicated
and one finds that it is easier to deal with the multidimensional
calculations by resorting to Monte Carlo integration techniques
[15]. Using the trial function (79), convergence is achieved much
more quickly for the $N=3,4$ cases. Comparing these results with
the $\beta=2$ calculation for $N=3,4$ is sufficient to give us an
idea of the effect of increasing the number of gluons. From Table
VII, for $c\overc$ and $N=3$, we can see that the $\beta=5$
results are better than the $\beta=4$ results, which are in turn
an improvement over $\beta=3$. However, none improve on the
answers for $\beta=2$. The same pattern holds for $N=4$ and for
$b\overb$.
\par\vfill\eject
\centerline {VI. CONCLUSIONS}
\par\medskip
Using the variational method and the Hamiltonian formalism of QCD,
we have derived an infinite chain of equations for a
quark-antiquark system interacting via an arbitrary number of
gluons. These coupled equations are in principle exact, but not
tractable. We attempt an approximation in which we decouple these
equations using an ansatz borrowed from QED-type calculations.
This leads to a sequence of (increasingly more complex) equations
for the $q\overq$ system with $\beta=0,1,2,...$ gluons present.
NonAbelian effects appear only for $\beta\ge 2$. For $\beta=0,1$
we have only the Abelian Coulomb-type interaction present, and
this leads to a Balmer-like mass-spectrum formula. We solve the
$\beta=2$ equation variationally, and work out predictions for the
low-lying energy levels of charmonium and bottomonium (at least in
the nonrelativistic limit). The results are encouraging since they
show a substantial improvement over the Abelian approximation. We
perform the same procedure for $\beta=3,4,5$ but find that the
results do not improve upon the ones for $\beta=2$. It would be
useful to see to what extent the results are improved if the
relativistic versions of these calculations are performed.

\centerline {ACKNOWLEDGMENTS}
\par\medskip
The support of the Natural Sciences and Engineering Research
Council of Canada for this project is gratefully acknowledged.
\par\vfill\eject
\centerline {APPENDIX A}
\par\medskip
In Equation (31), we use the following definitions: $$\eqalignno{
&{\tilde Q}_1(\beta,{\bf n},s_2,s,\lambda'',{\bf k})=
Q_1(\beta,{\bf n},s_2,s,0,{\bf k})+Q_1(\beta,{\bf n},s_2,s,3,{\bf
k})\cr &{\tilde Q}_2(\beta,{\bf p},s_1,s,\lambda'',{\bf k})=
Q_2(\beta,{\bf p},s_1,s,0,{\bf k})+Q_2(\beta,{\bf p},s_1,s,3,{\bf
k}) &(A1)\cr} $$ for $\lambda''=0$, and $$ \eqalignno{ &{\tilde
Q}_1(\beta,{\bf n},s_2,s,\lambda'',{\bf k})= Q_1(\beta,{\bf
n},s_2,s,\lambda'',{\bf k})\cr &{\tilde Q}_2(\beta,{\bf
p},s_1,s,\lambda'',{\bf k})= Q_2(\beta,{\bf
p},s_1,s,\lambda'',{\bf k})&(A2)\cr} $$ for $\lambda''=1,2$. And,
$$\eqalignno{ &{\tilde X} (\beta,k,j,\lambda''',\lambda',{\bf
k}')=X(\beta,k,j,0,0,{\bf k}')+X(\beta,k,j,3,3,{\bf k}')\cr
&\;\;\;\;\;\;\;\;\;\;\;\;\;\;\; +X(\beta,k,j,0,3,{\bf
k}')+X(\beta,k,j,3,0,{\bf k}') \;\;\;{\rm
for}\;\;\;\lambda',\lambda'''=0\cr &{\tilde X}
(\beta,k,j,\lambda''',\lambda',{\bf
k}')=X(\beta,k,j,0,\lambda',{\bf k}')+ X(\beta,k,j,3,\lambda',{\bf
k}')\;\;\;{\rm for}\;\;\;\lambda'=1,2;\lambda'''=0\cr &{\tilde X}
(\beta,k,j,\lambda''',\lambda',{\bf
k}')=X(\beta,k,j,\lambda''',0,{\bf k}')+
X(\beta,k,j,\lambda''',3,{\bf k}')\;\;\;{\rm
for}\;\;\;\lambda'=0;\lambda'''=1,2\cr &{\tilde X}
(\beta,k,j,\lambda''',\lambda',{\bf
k}')=X(\beta,k,j,\lambda''',\lambda',{\bf k}')\;\;\;{\rm
for}\;\;\;\lambda',\lambda'''=1,2\cr &&(A3)\cr} $$ where $$
Q_1(\beta,{\bf n},s_2,s,\lambda'',{\bf k})={v({\bf n},s_2)\;
\epslash^T_{\lambda''}({\bf k})\;{\bar v}({\bf n}-{\bf k},s)\over
\sqrt{\omega({\bf n})\;\omega({\bf n}-{\bf k})}}\;J_1^{\beta+1}
\eqno (A4) $$ $$ Q_2(\beta,{\bf p},s_1,s,\lambda'',{\bf k})={{\bar
u}({\bf p},s_1)\;\epslash_{\lambda''}({\bf k})\;u({\bf p}-{\bf
k},s)\over \sqrt{\omega({\bf p})\;\omega({\bf p}-{\bf
k})}}\;J_2^{\beta+1} \eqno (A5) $$ $$ Q_3(\beta,k,{\bf
n},s_2,s)={v({\bf n},s_2)\;\epslash_{\lambda_k}^T({\bf
q}_k)\;{\bar v}({\bf n}+{\bf q}_k,s)\over \sqrt{\omega({\bf
n})\;\omega({\bf n}+{\bf q}_k)}}\;J_3^{\beta-1}(z_k) \eqno (A6) $$
$$ Q_4(\beta,k,{\bf p},s_1,s)={{\bar u}({\bf
p},s_1)\;\epslash_{\lambda_k}({\bf q}_k)\;u({\bf p}+{\bf q}_k,s)
\over\sqrt{\omega({\bf p})\;\omega({\bf p}+{\bf
q}_k)}}\;J_4^{\beta-1}(z_k) \eqno (A7) $$ and $$\eqalignno{
X(\beta,k,j,\lambda''',\lambda',{\bf
k}')&=2\;K_6^{\beta}(z_k,z_j)\;\Biggl[
\fakebold{$\epsilon$}_{\lambda'}({\bf
k}')\cdot\fakebold{$\epsilon$}_{\lambda_k}({\bf q}_k)\;
\fakebold{$\epsilon$}_{\lambda_j}({\bf q}_j)\cdot
\fakebold{$\epsilon$}_{\lambda'''}({\bf q}_k+{\bf q}_j-{\bf
k}')\cr &-\fakebold{$\epsilon$}_{\lambda_k}({\bf
q}_k)\cdot\fakebold{$\epsilon$}_{\lambda_j}({\bf q}_j)\;
\fakebold{$\epsilon$}_{\lambda'}({\bf
k}')\cdot\fakebold{$\epsilon$}_{\lambda'''}({\bf q}_k+{\bf
q}_j-{\bf k}')\Biggr]&(A8)\cr} $$ And, lastly,(summing over
repeated indices), $$\eqalignno{ &I^{\beta}=
\{t^{z_{\beta}}_{l'_1\;{l'_2}}\cdots
t^{z_2}\;t^{z_1}_{l'_{\beta}\;l'_{\beta+1}}\}\;\delta_{l'_1\,j}
\delta_{l'_{\beta+1}\,i}\;\delta_{l_1\,i}\;\delta_{l_{\beta+1}\,
j}\cr &\;\;\;\;\;\times \beta!\;
\{t^{z_1}_{l_1\;l_2}\;t^{z_2}_{l_2\;l_3}\cdots
t^{z_{\beta}}_{l_{\beta}\;{l_{\beta+1}}}\}\cr
&J_1^{\beta+1}=\{t^{z_{\beta}}_{l'_1\;{l'_2}}\cdots
t^{z_2}\;t^{z_1}_{l'_{\beta}\;l'_{\beta+1}}\}\;t^a_{j\;i'}\;
\delta_{l_1\,i}\;\delta_{l_{\beta+2}\,j}\;\delta_{l'_1\,i'}\;
\delta_{l'_{\beta+1}\,i}\cr &\;\;\;\;\;\times (\beta+1)!\;
\{t^{a}_{l_1\;l_2}\;t^{z_1}_{l_2\;l_3}\cdots
t^{z_{\beta}}_{l_{\beta+1}\;{l_{\beta+2}}}\}\cr
&J_2^{\beta+1}=\{t^{z_{\beta}}_{l'_1\;{l'_2}}\cdots
t^{z_2}\;t^{z_1}_{l'_{\beta}\;l'_{\beta+1}}\}\;t^a_{i'\;i}\;
\delta_{l_1\,i}\;\delta_{l_{\beta+2}\,j}\;\delta_{l'_1\,j}\;
\delta_{l'_{\beta+1}\,i'}\cr &\;\;\;\;\;\times (\beta+1)!\;
\{t^{a}_{l_1\;l_2}\;t^{z_1}_{l_2\;l_3}\cdots
t^{z_{\beta}}_{l_{\beta+1}\;{l_{\beta+2}}}\}\cr
&J_3^{\beta-1}(z_k)=\{t^{z_{\beta}}_{l'_1\;{l'_2}}\cdots
t^{z_2}\;t^{z_1}_{l'_{\beta}\;l'_{\beta+1}}\}\;t^{z_k}_{j\;i'}\;
\delta_{l_1\,i}\;\delta_{l_{\beta}\,j}\;\delta_{l'_1\,i'}\;
\delta_{l'_{\beta+1}\,i}\cr &\;\;\;\;\;\times
(\beta-1)!\;\{t^{z_1}_{l_1\;l_2}\cdots{\tilde t^{z_k}}\cdots
t^{z_{\beta}}_{l_{\beta-1}\;{l_{\beta}}}\}\cr
&J_4^{\beta-1}(z_k)=\{t^{z_{\beta}}_{l'_1\;{l'_2}}\cdots
t^{z_2}\;t^{z_1}_{l'_{\beta}\;l'_{\beta+1}}\}\;t^{z_k}_{i'\;i}\;
\delta_{l_1\,i}\;
\delta_{l_{\beta}\,j}\;\delta_{l'_1\,j}\;\delta_{l'_{\beta+1}\,i'
}\cr &\;\;\;\;\;\times (\beta-1)!\;
\{t^{z_1}_{l_1\;l_2}\cdots{\tilde t^{z_k}}\cdots
t^{z_{\beta}}_{l_{\beta-1}\;{l_{\beta}}}\}\cr
&K_6^{\beta}(z_k,z_j)=f_{a\,z_j\,d}\;f_{a\,z_k\,e}\;
\{t^{z_{\beta}}_{l'_1\;{l'_2}}\cdots
t^{z_2}\;t^{z_1}_{l'_{\beta}\;l'_{\beta+1}}\}\;\delta_{l'_1\,j'}\
; \delta_{l'_{\beta+1}
\,i}\;\delta_{l_1\,i}\delta_{l_{\beta+1}\,j'}\cr &\;\;\;\;\;\times
\beta!\;\{t^{d}_{l_1\;l_2}\;t^{e}_{l_2\;l_3}\;t^{z_1}_{l_3\;l_4}
\cdots {\tilde {t^{z_k}}}\;{\tilde {t^{z_j}}}\cdots
t^{z_{\beta}}_{l_{\beta}\;{l_{\beta+1}}}\} & (A9)\cr} $$
\par
A closer look at our sums (A9) reveals the following
relationships: $$\eqalignno{ &J_1^{\beta+1}=J_2^{\beta+1}\cr
&J_3^{\beta}(z_{\beta+1})=J_4^{\beta}(z_{\beta+1})\cr
&I^{\beta+1}=(\beta+1)\;J_1^{\beta+1}=(\beta+1)^2\;J_3^{\beta}(z_
{\beta+1}) &(A10)\cr} $$ so that the expressions in (37) and (38)
(i.e., $\gamma(\beta)$) may be simplified to $$\eqalignno{
&\gamma(\beta)=\Biggl\{
{J_1^{\beta+1}\;J_4^{\beta}(z_{\beta+1})\over
I^{\beta}\;I^{\beta+1}}+
{J_2^{\beta+1}\;J_3^{\beta}(z_{\beta+1})\over
I^{\beta}\;I^{\beta+1}}
\Biggr\}=\Biggl\{{J_1^{\beta+1}\;J_3^{\beta}(z_{\beta+1})\over
I^{\beta}\;I^{\beta+1}}+
{J_2^{\beta+1}\;J_4^{\beta}(z_{\beta+1})\over
I^{\beta}\;I^{\beta+1}}\Biggr\}\cr
&=2\;{J_1^{\beta+1}\;J_4^{\beta}(z_{\beta+1})\over
I^{\beta}\;I^{\beta+1}}=2\;{I^{\beta+1}\over
(\beta+1)^3\;I^{\beta}}. &(A11)\cr} $$
\par\vfill\eject
\centerline {APPENDIX B}
\par\medskip
Removing the permutation terms in (29) leads us to the set of
equations: $$\eqalignno{ &\left[
\omega_{p}+\omega_{n}+m\;(m_0-m)\left({1\over \omega_{p}}+ {1\over
\omega_{n}}\right)+\sum_{k=1}^{\beta}\;|{\bf q}_k|-E
\right]\;I^{\beta}\;\phi^{\beta} ({\bf p},{\bf
n},s_1,s_2,q,\lambda)\cr
&+g'\;m\;\sum_{\lambda''=0}^2\sum_s\;\zeta_{\lambda''}\;\int d^3k
\;{{\tilde Q}_1(\beta,{\bf n},s_2,s,\lambda'',{\bf k})\over
\sqrt{|{\bf k}|}}\;\phi^{\beta+1}({\bf p},{\bf n}-{\bf
k},s_1,s,q,{\bf k},\lambda,\lambda'')\cr
&-g'\;m\;\sum_{\lambda''=0}^2\sum_s\;\zeta_{\lambda''}\;\int
d^3k\;{{\tilde Q}_2(\beta,{\bf p},s_1,s,\lambda'',{\bf k})\over
\sqrt{|{\bf k}|}}\;\phi^{\beta+1}({\bf p}-{\bf k},{\bf
n},s,s_2,q,{\bf k},\lambda,\lambda'')\cr
&+g'\;m\;\sum_s\;\sum_{k=1}^{\beta}\;{Q_3(\beta,k,{\bf
n},s_2,s)\over \sqrt{|{\bf q}_k|}}\;\phi^{\beta-1}({\bf p},{\bf
n}+{\bf q}_k,s_1,s,{\tilde q_k},{\tilde \lambda_k})\cr
&-g'\;m\;\sum_s\;\sum_{k=1}^{\beta}\;{Q_4(\beta,k,{\bf p},s_1,s)
\over\sqrt{|{\bf q}_k|}}\;\phi^{\beta-1}({\bf p}+{\bf q}_k,{\bf
n},s,s_2,{\tilde q_k},{\tilde\lambda_k})\cr &-{g'\over
2}\;\sum_{k\ne j \ne l}^{\beta}\;T_1(\beta,k,j,l)\;
\phi^{\beta-3}({\bf p},{\bf n},s_1,s_2, {\tilde q_k}\;{\tilde
q_j}\;{\tilde q_l},{\tilde \lambda_k}\; {\tilde
\lambda_j}\;{\tilde \lambda_l})\; {\delta^3({\bf q}_k+{\bf
q}_j+{\bf q}_l)\over \sqrt{|{\bf q}_k|\; |{\bf q}_j|\;|{\bf
q}_l|}} \cr &+{g'\over 2}\;\sum_{\lambda''=0}^2
\;\zeta_{\lambda''} \;\sum_{k \ne j}^{\beta} \;{{\tilde
T}_2(\beta,k,j,\lambda'')\;\phi^{\beta-1}({\bf p},{\bf n},s_1,s_2,
{\bf q}_k+{\bf q}_j, {\tilde {{\bf q}_k}}\; {\tilde {{\bf
q}_j}},\lambda'',{\tilde \lambda_k}\; {\tilde \lambda_j})\;
\over{\sqrt{|{\bf q}_k|\;|{\bf q}_j|\;|{\bf q}_k+{\bf q}_j|}}}\cr
&-{g'^2\over 8}\;\sum_{k\ne j \ne l \ne m}^{\beta}\;
F_1(\beta,k,j,l,m)\;\phi^{\beta-4}({\bf p},{\bf n},s_1,s_2,
{\tilde {{\bf q}_k}}\;{\tilde {{\bf q}_j}}\;{\tilde {{\bf q}_l}}\;
{\tilde {{\bf q}_m}},{\tilde {\lambda_j}}\;{\tilde {\lambda_k}}\;
{\tilde {\lambda_l}}\;{\tilde {\lambda_m}})\cr & \times
{\delta^3({\bf q}_k+{\bf q}_j +{\bf q}_l+{\bf q}_m)\over
\sqrt{|{\bf q}_k|\;|{\bf q}_j|\;|{\bf q}_l|\;|{\bf q}_m|}}\cr
&-{g'^2\over 8}\;\sum_{\lambda'=0}^2\;\zeta_{\lambda'}\;\sum_{k\ne
j \ne l}^{\beta}\;{{\tilde F}_2(\beta,k,j,l,\lambda')\;
\phi^{\beta-2}({\bf p},{\bf n},s_1,s_2,{\bf q}_k+{\bf q}_j+{\bf
q}_l,{\tilde {{\bf q}_k}}\;{\tilde {{\bf q}_j}}\; {\tilde {{\bf
q}_l}},\lambda',{\tilde {\lambda_k}}\;{\tilde {\lambda_j}}\;
{\tilde {\lambda_l}})\over \sqrt{|{\bf q}_k|\;|{\bf q}_j|\;|{\bf
q}_l|\;|{\bf q}_k+{\bf q}_j+ {\bf q}_l|}}\cr &-{g'^2\over
8}\;\sum_{\lambda',\lambda'''=0}^2\;\zeta_{\lambda'}\;
\zeta_{\lambda'''}\; \sum_{k\ne j}^{\beta}\;\int {d^3k' \over
\sqrt{|{\bf q}_k|\;|{\bf q}_j|\;|{\bf k}'|\; |{\bf q}_k+{\bf
q}_j-{\bf k}'|}}\cr &\times {\tilde X}
(\beta,k,j,\lambda''',\lambda',{\bf k}')\;\phi^{\beta}({\bf
p},{\bf n},s_1,s_2,{\bf q}_k+{\bf q}_j- {\bf k}',{\bf k}',{\tilde
{{\bf q}_k}}\; {\tilde {{\bf q}_j}},\lambda''',\lambda', {\tilde
{\lambda_k}}\;{\tilde {\lambda_j}})=0 & (B1)\cr} $$ where the
$Q$'s and ${\tilde X}$ are defined in APPENDIX A and $$\eqalignno{
&{\tilde T}_2(\beta,k,j,\lambda'')=
T_2(\beta,k,j,0)+T_2(\beta,k,j,3)\cr &{\tilde F}_2
(\beta,k,j,l,\lambda')=F_2(\beta,k,j,l,0)+F_2(\beta,k,j,l,3)&(B2)
\cr} $$ for $\lambda'',\lambda'=0$, and $$ \eqalignno{ &{\tilde
T}_2(\beta,k,j,\lambda'')= T_2(\beta,k,j,\lambda'')\cr &{\tilde
F}_2 (\beta,k,j,l,\lambda')=F_2(\beta,k,j,l,\lambda')&(B3)\cr} $$
for $\lambda'',\lambda'=1,2$.

Furthermore, $$
T_1(\beta,k,j,l)=G_2^{\beta-3}(z_k,z_j,z_l)\;\epsilon_{\lambda_k}
^{\nu}({\bf q}_k)\;\epsilon_{\nu\;\lambda_l} ({\bf
q}_l)\;\left[{\bf q}_k\cdot\fakebold{$\epsilon$}_{\lambda_j}({\bf
q}_j)- |{\bf q}_k|\; \epsilon_{\lambda_j}^0({\bf q}_j)\right]
\eqno (B4) $$ $$\eqalignno{
T_2(\beta,k,j,\lambda'')&=G_3^{\beta-1}(z_k,z_j)\cr & \times
\Biggl[ -\epsilon_{\lambda''}^{\nu}({\bf q}_k+{\bf q}_j)\;
\epsilon_{\nu\;\lambda_k}({\bf q}_k)\; \left[ ({\bf q}_k+{\bf
q}_j)\cdot\fakebold{$\epsilon$}_{\lambda_j}({\bf q}_j)- |{\bf
q}_k+{\bf q}_j|\; \epsilon_{\lambda_j}^0({\bf q}_j) \right]\cr
&+\epsilon_{\lambda_k}^{\nu}({\bf q}_k)\;\epsilon_{\nu\;\lambda_j}
({\bf q}_j)\;\left[ {\bf
q}_k\cdot\fakebold{$\epsilon$}_{\lambda''}({\bf q}_k+{\bf q}_j)-
|{\bf q}_k|\; \epsilon_{\lambda''}^0({\bf q}_k+{\bf q}_j)
\right]\cr &-\epsilon_{\lambda_k}^{\nu}({\bf
q}_k)\;\epsilon_{\nu\;\lambda''} ({\bf q}_k+{\bf q}_j)\;\left[
{\bf q}_k\cdot\fakebold{$\epsilon$}_{\lambda_j}({\bf q}_j)- |{\bf
q}_k|\; \epsilon_{\lambda_j}^0({\bf q}_j)
\right]\;\Biggr]&(B5)\cr} $$ $$
F_1(\beta,k,j,l,m)=K_2^{\beta-4}(z_k,z_j,z_l,z_m)\;
\epsilon_{\lambda_k \;\nu}({\bf
q}_k)\;\epsilon_{\lambda_j}^{\nu}({\bf q}_j)\left[
\epsilon^0_{\lambda_l}({\bf q}_l)\;\epsilon^0_{\lambda_m}({\bf
q}_m)+ \fakebold{$\epsilon$}_{\lambda_l}({\bf
q}_l)\cdot\fakebold{$\epsilon$}_{\lambda_m}({\bf q}_m)
\right]\eqno (B6) $$ $$\eqalignno{ &F_2(\beta,k,j,l,\lambda')=
K_3^{\beta-2}(z_k,z_j,z_l)\cr & \times
\Biggl[2\;\epsilon_{\lambda_k \;\nu}({\bf
q}_k)\;\epsilon_{\lambda'}^{\nu}({\bf q}_k+{\bf q}_j+ {\bf
q}_l)\left[\epsilon^0_{\lambda_j}({\bf q}_j)\;
\epsilon^0_{\lambda_l}({\bf q}_l)+
\fakebold{$\epsilon$}_{\lambda_j}({\bf
q}_j)\cdot\fakebold{$\epsilon$}_{\lambda_l}({\bf q}_l) \right]\cr
&+2\;\epsilon_{\lambda_j \;\nu}({\bf
q}_j)\;\epsilon_{\lambda_l}^{\nu}({\bf q}_l)\left[
\epsilon^0_{\lambda'}({\bf q}_k+{\bf q}_j+{\bf q}_l)\;\epsilon^0_{
\lambda_k}({\bf q}_k)+\fakebold{$\epsilon$}_{\lambda'}({\bf
q}_k+{\bf q}_j+ {\bf
q}_l)\cdot\fakebold{$\epsilon$}_{\lambda_k}({\bf q}_k)\right]
\Biggr].&(B7)\cr} $$ Finally, (summing over repeated indices),
$$\eqalignno{ &I^{\beta}= t^{z_{\beta}}_{l'_1\;{l'_2}}\cdots
t^{z_2}\;t^{z_1}_{l'_{\beta}\;l'_{\beta+1}}\;\delta_{l'_1\,j}\;
\delta_{l'_{\beta+1}\,i}\;\delta_{l_1\,i}\;\delta_{l_{\beta+1}\,j
} \cr &\;\;\;\;\;\times \left[
t^{z_1}_{l_1\;l_2}\;t^{z_2}_{l_2\;l_3}\cdots
t^{z_{\beta}}_{l_{\beta}\;{l_{\beta+1}}}+permutations[z_1,\cdots,
z_{\beta} ]\right]\cr
&J_1^{\beta+1}=t^{z_{\beta}}_{l'_1\;{l'_2}}\cdots
t^{z_2}\;t^{z_1}_{l'_{\beta}\;l'_{\beta+1}}\;t^a_{j\;i'}\;
\delta_{l_1\,i}\;\delta_{l_{\beta+2}\,j}\;\delta_{l'_1\,i'}\;
\delta_{l'_{\beta+1}\,i}\cr &\;\;\;\;\;\times \left[
t^{a}_{l_1\;l_2}\;t^{z_1}_{l_2\;l_3}\cdots
t^{z_{\beta}}_{l_{\beta+1}\;{l_{\beta+2}}}+permutations[a,z_1,
\ldots,z_{\beta} ] \right]\cr
&J_2^{\beta+1}=t^{z_{\beta}}_{l'_1\;{l'_2}}\cdots
t^{z_2}\;t^{z_1}_{l'_{\beta}\;l'_{\beta+1}}\;\;t^a_{i'\;i}\;
\delta_{l_1\,i}\;\delta_{l_{\beta+2}\,j}\;\delta_{l'_1\,j}\;
\delta_{l'_{\beta+1}\,i'}\cr &\;\;\;\;\;\times \left[
t^{a}_{l_1\;l_2}\;t^{z_1}_{l_2\;l_3}\cdots
t^{z_{\beta}}_{l_{\beta+1}\;{l_{\beta+2}}}+permutations[a,z_1,
\ldots,z_{\beta}]  \right]\cr
&J_3^{\beta-1}(z_k)=t^{z_{\beta}}_{l'_1\;{l'_2}}\cdots
t^{z_2}\;t^{z_1}_{l'_{\beta}\;l'_{\beta+1}}\;t^{z_k}_{j\;i'}\;
\delta_{l_1\,i}\;\delta_{l_{\beta}\,j}\;\delta_{l'_1\,i'}\;
\delta_{l'_{\beta+1}\,i}\cr &\;\;\;\;\;\times \left[
t^{z_1}_{l_1\;l_2}\cdots{\tilde t^{z_k}}\cdots
t^{z_{\beta}}_{l_{\beta-1}\;{l_{\beta}}}+permutations[z_1,\ldots,
{\tilde z_k},\ldots,z_{\beta} ] \right]\cr
&J_4^{\beta-1}(z_k)=t^{z_{\beta}}_{l'_1\;{l'_2}}\cdots
t^{z_2}\;t^{z_1}_{l'_{\beta}\;l'_{\beta+1}}\;t^{z_k}_{i'\;i}\;
\delta_{l_1\,i}\;
\delta_{l_{\beta}\,j}\;\delta_{l'_1\,j}\;\delta_{l'_{\beta+1}\,i'
}\cr &\;\;\;\;\;\times \left[  t^{z_1}_{l_1\;l_2}\cdots{\tilde
t^{z_k}}\cdots
t^{z_{\beta}}_{l_{\beta-1}\;{l_{\beta}}}+permutations[z_1,\ldots,
{\tilde z_k},\ldots,z_{\beta} ] \right]\cr
&G_2^{\beta-3}(z_k,z_j,z_l)=  i\;f_{z_k\,z_j\,z_l}\;
t^{z_{\beta}}_{l'_1\;{l'_2}}\cdots
t^{z_2}\;t^{z_1}_{l'_{\beta}\;l'_{\beta+1}}\;\delta_{l'_1\,j'}\;
\delta_{l'_{\beta+1}\,i}\;\delta_{l_1\,i}\;\delta_{l_{\beta-2}\,j
'}\cr &\;\;\;\;\;\times \left[ t^{z_1}_{l_1\;l_2}\cdots {\tilde
t^{z_k}}\;{\tilde t^{z_j}}\;{\tilde t^{z_l}}\cdots
t^{z_{\beta}}_{l_{\beta-3}\;{l_{\beta-2}}}+permutations[z_1,
\ldots, {\tilde z_k},{\tilde z_j},{\tilde
z_l},\ldots,z_{\beta}]\right]\cr
&G_3^{\beta-1}(z_k,z_j)=i\;f_{z_k\,z_j\,c}
\;t^{z_{\beta}}_{l'_1\;{l'_2}}\cdots
t^{z_2}\;t^{z_1}_{l'_{\beta}\;l'_{\beta+1}}\;\delta_{l'_1\,j'}\;
\delta_{l'_{\beta+1}\,i}\;\delta_{l_1\,i}\;\delta_{l_{\beta}\,j'}
\cr &\;\;\;\;\;\times
\left[t^{c}_{l_1\;l_2}\;t^{z_1}_{l_2\;l_3}\cdots {\tilde
t^{z_k}}\;{\tilde t^{z_j}}\cdots
t^{z_{\beta}}_{l_{\beta-1}\;{l_{\beta}}}+permutations
[c,z_1,\ldots,{\tilde z_k},{\tilde z_j},\ldots,z_{\beta}]
\right]\cr
&K_2^{\beta-4}(z_k,z_j,z_l,z_m)=f_{a\,z_l\,z_k}\;f_{a\,z_m
\,z_j}\;t^{z_{\beta}}_{l'_1\;{l'_2}}\cdots
t^{z_2}\;t^{z_1}_{l'_{\beta}\;l'_{\beta+1}}\;\;\delta_{l'_1\,j'}\
; \delta_{l'_{\beta+1}
\,i}\;\delta_{l_1\,i}\delta_{l_{\beta-3}\,j'}\cr &\;\;\;\;\;\times
\left[t^{z_1}_{l_1\;l_2}\cdots {\tilde t^{z_k}}\;{\tilde
t^{z_j}}\;{\tilde t^{z_l}}\;{\tilde t^{z_m}}\cdots
t^{z_{\beta}}_{l_{\beta-4}\;{l_{\beta-3}}}
+permutations[z_1,\ldots,{\tilde z_k},{\tilde z_j},{\tilde z_l},
{\tilde z_m},\ldots,z_{\beta}]\right]\cr
&K_3^{\beta-2}(z_k,z_j,z_l)=f_{a\,z_j\,z_k}\;f_{a\,z_l\,e}\;t^{z_
{\beta}}_{l'_1\;{l'_2}}\cdots
t^{z_2}\;t^{z_1}_{l'_{\beta}\;l'_{\beta+1}}\;\delta_{l'_1\,j'}\;
\delta_{l'_{\beta+1}
\,i}\;\delta_{l_1\,i}\delta_{l_{\beta-1}\,j'}\cr
&\;\;\;\;\;\times\left[t^{e}_{l_1\;l_2}\;
t^{z_1}_{l_2\;l_3}\cdots{\tilde {t^{z_k}}}\;{\tilde {t^{z_j}}}\;
{\tilde {t^{z_l}}}\cdots
t^{z_{\beta}}_{l_{\beta-2}\;{l_{\beta-1}}}+permutations[e,z_1,
\ldots, {\tilde {z_k}},{\tilde {z_j}},{\tilde
{z_l}},\ldots,z_{\beta}]\right]\cr
&K_6^{\beta}(z_k,z_j)=f_{a\,z_j\,d}\;f_{a\,z_k\,e}\;
t^{z_{\beta}}_{l'_1\;{l'_2}}\cdots
t^{z_2}\;t^{z_1}_{l'_{\beta}\;l'_{\beta+1}}\;\delta_{l'_1\,j'}\;
\delta_{l'_{\beta+1}
\,i}\;\delta_{l_1\,i}\delta_{l_{\beta+1}\,j'}\cr &\;\;\;\;\;\times
\left[
t^{d}_{l_1\;l_2}\;t^{e}_{l_2\;l_3}\;t^{z_1}_{l_3\;l_4}\cdots
{\tilde {t^{z_k}}}\;{\tilde {t^{z_j}}}\cdots
t^{z_{\beta}}_{l_{\beta}\;{l_{\beta+1}}}
+permutations[d,e,z_1,\ldots,{\tilde {z_k}}\; {\tilde
{z_j}},\ldots,z_{\beta}]\right]\cr & &(B8)\cr} $$
\par\vfill\eject
\centerline {References}
\par\medskip
\noindent [1] W. Dykshoorn, R. Koniuk and J. Darewych, in {\it
Variational Calculations in Quantum Field Theory}, edited by L.
Polley and DEL Pottinger (Singapore, World Scientific, 1988) pp.
188-192.
\par
\noindent [2] W. Dykshoorn and R. Koniuk, Phys. Rev. A {\bf 41},
64 (1990).
\par
\noindent [3] J.W. Darewych and M. Horbatsch, J. Phys. B {\bf 22},
973 (1989) and {\bf 23}, 337 (1990).
\par
\noindent [4] T. Zhiang, L. Xiao and R. Koniuk, Can. J. Phys. {\bf
70}, 670 (1992).
\par
\noindent [5] J. W. Darewych, M. Horbatsch and R. Koniuk, Phys.
Rev. D {\bf 45}, 675 (1992).
\par
\noindent [6] W.C. Berseth and J.W. Darewych, Phys. Lett. A {\bf
178}, 347 (1993) and Erratum {\bf 185}, 503 (1994).
\par
\noindent [7] L. Di Leo and J.W. Darewych, Can. J. Phys. {\bf 70},
412 (1992) and {\bf 71}, 365 (1993).
\par
\noindent [8] T. Zhang and R. Koniuk, Phys. Rev. D {\bf 43}, 1688
(1991) and {\bf 48}, 5382 (1993).
\par
\noindent [9] J.W. Darewych, Ukr. J. Phys {\bf 41}, 41 (1996).
\par
\noindent [10] J.R. Spence and J.P. Vary, Phys. Rev. C {\bf 52},
1668 (1995).
\par
\noindent [11] Review of Particle Properties, 1998, European
Physical Journal.\par \noindent [12] F. Mandl and G. Shaw, {\it
Quantum Field Theory}, (John Wiley $\&$ Sons, Cichester, 1984).
\par
\noindent [13] J.W. Darewych {\it et al}, Phys Rev C {\bf 47},
1885 (1993).
\par
\noindent [14] J.D. Bjorken and S.D. Drell, {\it Relativistic
Quantum Fields}, (McGraw-Hill, New York, 1965).
\par
\noindent [15]  W.H. Press, S.A. Teukolsky, W.T. Vetterling, and
B.P. Flannery, {\it Numerical Recipes}. (Cambridge University
Press, 1992) pp. 309-314.

\par\vfill\eject
Table I. Predicted energies for $c\overc$ (nonAbelian terms
suppressed)
\par\medskip
\vbox{\offinterlineskip \hrule \halign{&\vrule#&
  \strut\quad\hfil#\quad\cr
height2pt&\omit&&\omit&&\omit&&\omit&&\omit&&\omit&\cr
&$n$\hfil&&$\beta=0$\hfil&&$\beta=1$\hfil&&$\beta=2$\hfil&&$Exp't
(MeV)$\hfil&\cr
height2pt&\omit&&\omit&&\omit&&\omit&&\omit&&\omit&&\omit&\cr
\noalign{\hrule}
height2pt&\omit&&\omit&&\omit&&\omit&&\omit&&\omit&\cr
&$1$&&$\alpha=0.9542$&&$\alpha=2.181$&&$\alpha=2.672$&&$3096.88$&
\cr &$2$&&$m_c=1941$&&$m_c=1941$&&$m_c=1941$&&$3686$&\cr
height2pt&\omit&&\omit&&\omit&&\omit&&\omit&&\omit&\cr
\noalign{\hrule}
height2pt&\omit&&\omit&&\omit&&\omit&&\omit&&\omit&\cr
&$3$&&$3795$&&$3795$&&$3795$&&$4040$&\cr
&$4$&&$3833$&&$3833$&&$3833$&&$4415$&\cr
&$5$&&$3851$&&$3851$&&$3851$&&$----$&\cr
&$6$&&$3860$&&$3860$&&$3860$&&$----$&\cr
height2pt&\omit&&\omit&&\omit&&\omit&&\omit&&\omit&\cr} \hrule}
\par\medskip
Table II. Predicted energies for $b\overb$ (nonAbelian terms
suppressed)
\par\medskip
\vbox{\offinterlineskip \hrule \halign{&\vrule#&
  \strut\quad\hfil#\quad\cr
height2pt&\omit&&\omit&&\omit&&\omit&&\omit&&\omit&\cr
&$n$\hfil&&$\beta=0$\hfil&&$\beta=1$\hfil&&$\beta=2$\hfil&&$Exp't
(GeV)$\hfil&\cr
height2pt&\omit&&\omit&&\omit&&\omit&&\omit&&\omit&&\omit&\cr
\noalign{\hrule}
height2pt&\omit&&\omit&&\omit&&\omit&&\omit&&\omit&\cr
&$1$&&$\alpha=0.575135$&&$\alpha=1.31459$&&$\alpha=1.61038$&&$9.4
6037$&\cr
&$2$&&$m_b=5.10547$&&$m_b=5.10547$&&$m_b=5.10547$&&$10.02330$&\cr
height2pt&\omit&&\omit&&\omit&&\omit&&\omit&&\omit&\cr
\noalign{\hrule}
height2pt&\omit&&\omit&&\omit&&\omit&&\omit&&\omit&\cr
&$3$&&$10.1275$&&$10.1275$&&$10.1275$&&$10.3553$&\cr
&$4$&&$10.1640$&&$10.1640$&&$10.1640$&&$10.5800$&\cr
&$5$&&$10.1809$&&$10.1809$&&$10.1809$&&$10.865$&\cr
&$6$&&$10.1901$&&$10.1901$&&$10.1901$&&$11.019$&\cr
height2pt&\omit&&\omit&&\omit&&\omit&&\omit&&\omit&\cr} \hrule}
\par\medskip
Table III. Predicted energies for $c\overc$ (nonAbelian terms
present, $\beta=2$) with trial function (79)
\par\medskip
\vbox{\offinterlineskip \hrule \halign{&\vrule#&
  \strut\quad\hfil#\quad\cr
height2pt&\omit&&\omit&&\omit&&\omit&&\omit&&\omit&\cr
&$n$\hfil&&$N=1.75$\hfil&&$N=2$\hfil&&$N=3$\hfil&&$Exp't
(MeV)$\hfil&\cr
height2pt&\omit&&\omit&&\omit&&\omit&&\omit&&\omit&&\omit&\cr
\noalign{\hrule}
height2pt&\omit&&\omit&&\omit&&\omit&&\omit&&\omit&\cr
&$1$&&$\alpha=.7372$&&$\alpha=1.053$&&$\alpha=1.550$&&$3096.88$&
\cr &$2$&&$m_c=3370$&&$m_c=2717$&&$m_c=2267$&&$3686$&\cr
height2pt&\omit&&\omit&&\omit&&\omit&&\omit&&\omit&\cr
\noalign{\hrule}
height2pt&\omit&&\omit&&\omit&&\omit&&\omit&&\omit&\cr
&$3$&&$3871$&&$3863$&&$3848$&&$4040$&\cr
&$4$&&$3961$&&$3948$&&$3923$&&$4415$&\cr
&$5$&&$4014$&&$3998$&&$3965$&&$----$&\cr
&$6$&&$4050$&&$4031$&&$3993$&&$----$&\cr
&$\infty$&&$4223$&&$4190$&&$4121$&&$----$&\cr
height2pt&\omit&&\omit&&\omit&&\omit&&\omit&&\omit&\cr} \hrule}
\par\medskip
Table IV. Predicted energies for $b\overb$ (nonAbelian terms
present, $\beta=2$) with trial function (79)
\par\medskip
\vbox{\offinterlineskip \hrule \halign{&\vrule#&
  \strut\quad\hfil#\quad\cr
height2pt&\omit&&\omit&&\omit&&\omit&&\omit&&\omit&\cr
&$n$\hfil&&$N=1.75$\hfil&&$N=2$\hfil&&$N=3$\hfil&&$Exp't
(GeV)$\hfil&\cr
height2pt&\omit&&\omit&&\omit&&\omit&&\omit&&\omit&&\omit&\cr
\noalign{\hrule}
height2pt&\omit&&\omit&&\omit&&\omit&&\omit&&\omit&\cr
&$1$&&$\alpha=0.5200$&&$\alpha=.7016$&&$\alpha=.9804$&&$9.46037$&
\cr &$2$&&$m_b=6.4710$&&$m_b=5.847$&&$m_b=5.4169$&&$10.02330$&\cr
height2pt&\omit&&\omit&&\omit&&\omit&&\omit&&\omit&\cr
\noalign{\hrule}
height2pt&\omit&&\omit&&\omit&&\omit&&\omit&&\omit&\cr
&$3$&&$10.1999$&&$10.1928$&&$10.1781$&&$10.3553$&\cr
&$4$&&$10.2862$&&$10.2742$&&$10.2494$&&$10.5800$&\cr
&$5$&&$10.3372$&&$10.3219$&&$10.2902$&&$10.865$&\cr
&$6$&&$10.3710$&&$10.3533$&&$10.3166$&&$11.019$&\cr
&$\infty$&&$10.5366$&&$10.5046$&&$10.4386$&&$----$&\cr
height2pt&\omit&&\omit&&\omit&&\omit&&\omit&&\omit&\cr} \hrule}
\par\medskip
\par\vfill\eject
Table V. Predicted energies for $c\overc$ (nonAbelian terms
present, $\beta=2$) with trial function (80)
\par\medskip
\vbox{\offinterlineskip \hrule \halign{&\vrule#&
  \strut\quad\hfil#\quad\cr
height2pt&\omit&&\omit&&\omit&&\omit&&\omit&&\omit&\cr
&$n$\hfil&&$N=.5$\hfil&&$N=1$\hfil&&$N=2$\hfil&&$Exp't
(MeV)$\hfil&\cr
height2pt&\omit&&\omit&&\omit&&\omit&&\omit&&\omit&&\omit&\cr
\noalign{\hrule}
height2pt&\omit&&\omit&&\omit&&\omit&&\omit&&\omit&\cr
&$1$&&$\alpha=.1014$&&$\alpha=.9197$&&$\alpha=1.592$&&$3096.88$&
\cr &$2$&&$m_c=18119$&&$m_c=2934$&&$m_c=2244$&&$3686$&\cr
height2pt&\omit&&\omit&&\omit&&\omit&&\omit&&\omit&\cr
\noalign{\hrule}
height2pt&\omit&&\omit&&\omit&&\omit&&\omit&&\omit&\cr
&$3$&&$3881$&&$3867$&&$3846$&&$4040$&\cr
&$4$&&$3979$&&$3954$&&$3920$&&$4415$&\cr
&$5$&&$4037$&&$4006$&&$3962$&&$----$&\cr
&$6$&&$4076$&&$4040$&&$3989$&&$----$&\cr
&$\infty$&&$4270$&&$4205$&&$4114$&&$----$&\cr
height2pt&\omit&&\omit&&\omit&&\omit&&\omit&&\omit&\cr} \hrule}
\par\medskip
Table VI. Predicted energies for $b\overb$ (nonAbelian terms
present, $\beta=2$) with trial function (80)
\par\medskip
\vbox{\offinterlineskip \hrule \halign{&\vrule#&
  \strut\quad\hfil#\quad\cr
height2pt&\omit&&\omit&&\omit&&\omit&&\omit&&\omit&\cr
&$n$\hfil&&$N=.5$\hfil&&$N=1$\hfil&&$N=2$\hfil&&$Exp't
(GeV)$\hfil&\cr
height2pt&\omit&&\omit&&\omit&&\omit&&\omit&&\omit&&\omit&\cr
\noalign{\hrule}
height2pt&\omit&&\omit&&\omit&&\omit&&\omit&&\omit&\cr
&$1$&&$\alpha=0.0931$&&$\alpha=.6258$&&$\alpha=1.0037$&&$9.46037$
&\cr &$2$&&$m_b=20.56$&&$m_b=6.0538$&&$m_b=5.3948$&&$10.02330$&\cr
height2pt&\omit&&\omit&&\omit&&\omit&&\omit&&\omit&\cr
\noalign{\hrule}
height2pt&\omit&&\omit&&\omit&&\omit&&\omit&&\omit&\cr
&$3$&&$10.2098$&&$10.1960$&&$10.1767$&&$10.3553$&\cr
&$4$&&$10.3029$&&$10.2796$&&$10.2470$&&$10.5800$&\cr
&$5$&&$10.3586$&&$10.3288$&&$10.2871$&&$10.865$&\cr
&$6$&&$10.3958$&&$10.3612$&&$10.3130$&&$11.019$&\cr
&$\infty$&&$10.5812$&&$10.5190$&&$10.4322$&&$----$&\cr
height2pt&\omit&&\omit&&\omit&&\omit&&\omit&&\omit&\cr} \hrule}
\par\medskip
\par\medskip
Table VII. Predicted energies for $c\overc$ (nonAbelian terms
present, $\beta=2,3,4,5$) with $N=3$ and trial function (79)
\par\medskip
\vbox{\offinterlineskip \hrule \halign{&\vrule#&
  \strut\quad\hfil#\quad\cr
height2pt&\omit&&\omit&&\omit&&\omit&&\omit&&\omit&&\omit&\cr
&$n$\hfil&&$\beta=5$\hfil&&$\beta=4$\hfil&&$\beta=3$\hfil&&$\beta=2$\hfil&&
$Ex(MeV)$\hfil&\cr
height2pt&\omit&&\omit&&\omit&&\omit&&\omit&&\omit&&\omit&&\omit&\cr
\noalign{\hrule}
height2pt&\omit&&\omit&&\omit&&\omit&&\omit&&\omit&&\omit&\cr
&$1$&&$\alpha=1.886-1.883
$&&$\alpha=1.930-1.928$&&$\alpha=1.799-1.798$&&$\alpha=1.550$&&$3096.88$&
\cr
&$2$&&$m_c=2179-2180$&&$m_c=2155-2156$&&$m_c=2129$&&$m_c=2267$&&$3686$&\cr
height2pt&\omit&&\omit&&\omit&&\omit&&\omit&&\omit&&\omit&\cr
\noalign{\hrule}
height2pt&\omit&&\omit&&\omit&&\omit&&\omit&&\omit&&\omit&\cr
&$3$&&$3841-3842$&&$3839$&&$3836$&&$3848$&&$4040$&\cr
&$4$&&$3912$&&$3908$&&$3903$&&$3923$&&$4415$&\cr
&$5$&&$3951$&&$3946$&&$3940$&&$3965$&&$---$&\cr
&$6$&&$3977$&&$3971$&&$3964$&&$3993$&&$---$&\cr
&$\infty$&&$4091-4092$&&$4081$&&$4069$&&$4121$&&$---$&\cr
height2pt&\omit&&\omit&&\omit&&\omit&&\omit&&\omit&&\omit&&\omit&\cr}
\hrule}
\par\vfill\eject
\par\vfill\eject
\bye